\documentclass{emulateapj}
\usepackage{natbib}
\usepackage{graphicx}
\usepackage{epstopdf}
\usepackage{color}

\newcommand{\kms}{km s$^{-1}$} 
\newcommand{\Ha}{H$\alpha$}
\newcommand{\Hb}{H$\beta$}
\newcommand{\ha}{H$\alpha$}
\newcommand{\hb}{H$\beta$}
\newcommand{\oiii}{[OIII]$\lambda$5007}
\newcommand{\oiiit}{[OIII]$\lambda$4363}
\newcommand{\oii}{[OII]$\lambda$3727}

\newcommand{\lya}{Ly$\alpha$}
\newcommand{\LyA}{\lya}
\newcommand{\fesc}{$f^{Ly\alpha}_{esc}$}

\def\arcsec{\hbox{$^{\prime\prime}$}}

\def\sun{\hbox{$\odot$}}
\def\earth{\hbox{$\oplus$}}
\def\lesssim{\mathrel{\hbox{\rlap{\hbox{%
 \lower4pt\hbox{$\sim$}}}\hbox{$<$}}}}
\def\gtrsim{\mathrel{\hbox{\rlap{\hbox{%
 \lower4pt\hbox{$\sim$}}}\hbox{$>$}}}}

\begin{document}
\title{\lya\ profile, dust, and prediction of \lya\ escape fraction in Green Pea Galaxies}

\author{Huan Yang\altaffilmark{1,2}, Sangeeta Malhotra\altaffilmark{2,3}, Max Gronke\altaffilmark{4}, James E. Rhoads\altaffilmark{2,3}, Claus Leitherer\altaffilmark{5}, Aida Wofford\altaffilmark{6}, Tianxing Jiang\altaffilmark{2}, Mark Dijkstra\altaffilmark{4}, V. Tilvi\altaffilmark{2}, Junxian Wang\altaffilmark{1}}

\altaffiltext{1}{CAS Key Laboratory for Research in Galaxies and Cosmology, Department of Astronomy, University of Science and Technology of China; huan.y@asu.edu}
\altaffiltext{2}{Arizona State University, School of Earth and Space Exploration}
\altaffiltext{3}{NASA Goddard Space Flight Center}
\altaffiltext{4}{Institute of Theoretical Astrophysics, University of Oslo, Norway}
\altaffiltext{5}{Space Telescope Science Institute}
\altaffiltext{6}{National Autonomous University of Mexico, Institute of Astronomy}

\begin{abstract}

We studied Lyman-$\alpha$ (\lya) escape in a statistical sample of 43 Green Peas with HST/COS \lya\ spectra. Green Peas are nearby star-forming galaxies with strong \oiii\ emission lines. Our sample is four times larger than the previous sample and covers a much more complete range of Green Pea properties. We found that about 2/3 of Green Peas are strong \lya\ line emitters with rest-frame \lya\ equivalent width $>20$~\AA. The \lya\ profiles of Green Peas are diverse. The \lya\ escape fraction, defined as the ratio of observed \lya\ flux to intrinsic \lya\ flux, shows anti-correlations with a few \lya\ kinematic features -- both the blue peak and red peak velocities, the peak separations, and FWHM of the red portion of the \lya\ profile. Using properties measured from SDSS optical spectra, we found many correlations -- \lya\ escape fraction generally increases at lower dust reddening, lower metallicity, lower stellar mass, and higher [OIII]/[OII] ratio. We fit their \lya\ profiles with the HI shell radiative transfer model and found \lya\ escape fraction anti-correlates with the best-fit $N_{HI}$. Finally, we fit an empirical linear relation to predict \fesc\ from the dust extinction and \lya\ red peak velocity. The standard deviation of this relation is about 0.3 dex. This relation can be used to isolate the effect of IGM scatterings from \lya\ escape and to probe the IGM optical depth along the line of sight of each $z>7$ \lya\ emission line galaxy in the JWST era.

\end{abstract}

\section{Introduction}

In young star forming galaxies, Lyman continuum (LyC) photons from hot stars ionize the surrounding hydrogen gas, and \lya\ photons come from the recombination of hydrogen gas. 
The \lya\ emission line is a powerful tool in discovering and studying high redshift galaxies. Thousands of high redshift \lya\ emission line galaxies (LAE) have been found in the last two decades (e.g. Dey et al. 1998; Hu et al. 1998; Rhoads et al. 2000; Ouchi et al. 2003; Gawiser et al. 2006; Wang et al. 2009; Kashikawa et al. 2011; Erb et al. 2014; Matthee et al. 2014; Zheng et al. 2016). 
These high redshift LAEs generally have small size, low stellar mass, low dust extinction, low metallicity, young age, and high specific star formation rate (sSFR) (e.g. Malhotra 2012; Bond et al. 2010; Gawiser et al. 2007; Pirzkal et al. 2007; Finkelstein et al. 2008). 
At $2\lesssim z \lesssim 6$, these LAEs are an important population of star-forming galaxies, and they constitute an increasing fraction of Lyman break galaxies across that range, reaching $\sim60\%$ of Lyman break galaxies (LBGs) at redshift $z\sim$6 (Stark et al. 2011). 

A current frontier is searching for LAEs in the epoch of Cosmic Reionization.  As \lya\ photons propagate from a LAE to the observer, they pass through the intergalactic medium (IGM) and will be scattered away from the line of sight by HI in IGM. So \lya\ line can be used to probe reionization of IGM (e.g. Malhotra \& Rhoads 2004; Treu et al. 2012; Pentericci et al. 2014; Tilvi et al. 2014; Matthee et al. 2015; Santos et al. 2016). These \lya\ based methods can effectively probe HI fraction in the later half of reionization. One major goal of JWST is to observe the \lya\ and rest-frame optical lines spectra of $z>7$ galaxies and probe reionization with \lya\ lines. However, the challenge is to isolate the impact of IGM from other effects that may diminish \lya. The \lya\ photons have to escape out of the galaxies before passing through the IGM and being observed, i.e. $(Observed\ Ly\alpha)=(Intrinsic\ Ly\alpha)\times\ (Ly\alpha\ escape\ fraction) \times (IGM\ Transmission)$. 
The \lya\ escape fraction describes how many \lya\ photons escape out of both interstellar medium (ISM) and circum-galactic medium (CGM) of a LAE. Thus, to use \lya\ reionization tests, we have to understand \lya\ escape and predict \lya\ escape fraction from other properties. 

\lya\ escape is also related to the LyC escape process. A large fraction (~9/12) of known LyC leakers are LAEs (Leitet et al. 2013; Borthakur et al. 2014; Izotov et al. 2016; Leitherer et al. 2016; de Barros et al. 2016; Shapley et al. 2016).  LAEs at the reionization epoch may be major contributors of ionizing photons.  \lya\ line profiles may be used as a tool for detecting LyC leakers (Verhamme et al. 2015; Alexandroff et al. 2015; Dijkstra et al. 2016).  Understanding \lya\ escape is very useful for the study of LyC escape.

As \lya\ is a resonance line, it has a high cross-section for HI scattering. The emergent \lya\ emission has a complicatedly dependence on the amount of dust, the HI gas column density ($N_{HI}$), the kinematics of HI gas, and the geometric distribution of HI gas and dust (e.g. Neufeld 1990; Charlot \& Fall 1993; Ahn et al. 2001; Verhamme et al. 2006; Dijkstra et al. 2006; Laursen et al. 2013). The scattering of \lya\ photons can significantly modify the \lya\ line profile. LAEs usually show asymmetric or a double-peaked \lya\ emission line profiles (e.g. Rhoads et al 2003; Kashikawa et al. 2011; Erb et al. 2014). Therefore the \lya\ line profile carries a lot of information about the resonant scatterings and can be used to probe the HI gas properties.

To study \lya\ escape, it is ideal to have a large sample of LAEs and measure high quality \lya\ line spectrum, many optical emission lines, HI gas properties, and multiple other galactic properties. So we can test what properties make \lya\ escape, and finally {\it predict} \lya\ escape fraction from those properties. At high redshift, however, absorption by the intergalactic \lya\ forest prevents reliable measurements of the blue portion of \lya\ emission lines.  Other crucial observations are also impractical, both because high-$z$ LAEs are faint, and because some features (notably rest-optical emission lines) are redshifted to $\lambda_{obs} > 2.4{\mu}m$, where presently available instruments lack sensitivity.
Therefore many studies seek to solve the \lya\ escape problem by observing low-$z$ galaxies with similar properties to high-$z$ LAEs (e.g. Giavalisco et al. 1996; Kunth et al. 1998; Mas-Hesse et al. 2003; Deharveng et al. 2008; Finkelstein et al. 2009; Atek et al. 2009; Leitherer et al. 2011; Heckman et al. 2011; Cowie et al. 2011; Wofford et al. 2013; Hayes et al. 2005, 2014; Ostlin et al. 2014; Rivera-Thorsen et al. 2015). However, low-$z$ LAEs are rare and many nearby \lya\ emission line galaxies are older and more evolved galaxies than typical high-$z$ LAEs and may be a different population of \lya\ emitters. Perhaps the most relevant nearby analogs of high-$z$ LAEs are Green Pea galaxies (Jaskot \& Oey 2014; Henry et al. 2015; Yang et al. 2016a, hereafter Paper I). 

Green Pea galaxies were discovered in the citizen science project Galaxy Zoo, in which public volunteers morphologically classified millions of galaxies from the Sloan Digital Sky Survey (SDSS). Green Peas are compact galaxies that are unresolved in SDSS images. The green color is because the [OIII] doublet dominates the flux of SDSS $r$-band which is mapped to the green channel in the SDSS's false-color {\it gri}-band images (Lupton et al. 2004). They generally have small stellar masses ($\sim 10^{8-10} M_{\odot}$), low metallicities for their stellar masses, high specific star formation rates (sSFR), and large \oiii/\oii\ (hereafter [OIII]/[OII]) ratio (Cardamone et al. 2009; Amorin et al. 2010; Izotov et al. 2011). The UV spectra of 17 Green Peas generally show strong Ly$\alpha$ emission lines (Paper I; Jaskot \& Oey 2014; Henry et al. 2015; Izotov et al. 2016; Verhamme et al. 2016). These studies have explored the relation of \fesc\ and dust, metallicity, \lya\ profiles, and metal absorption lines with small samples of Green Peas. Besides the small sample size, the previous samples of Green Peas tend to be lower metallicity and lower dust extinction than the whole Green Pea sample. In our HST program, we observed an additional 20 Green Peas in order to have a statistical sample that spans a range of galaxy properties such as metallicity, dust extinction, and star-formation rate (SFR).

In this paper, we use HST/COS \lya\ spectra of Green Peas to study the mechanism of \lya\ escape. In Section 2, we show the sample and observations. In Section 3, we describe the measurement and properties of \lya\ equivalent width and escape fraction. In Section 4, we show the relation between \lya\ escape and \lya\ kinematic features. In Section 5, we show the relation between \lya\ escape and dust extinction, metallicity, stellar mass, morphology, and [OIII]/[OII] ratio. In Section 6, we fit the \lya\ profiles with radiative transfer model. In Section 7, we show an empirical relation to predict \lya\ escape fraction and discuss its applications on probing reionization.

\section{Sample and Observations}
\subsection{The Sample}
Since the strong \oiii\ line makes Green Pea galaxies have special optical broadband colors, we can select a few thousand Green Pea candidates from the SDSS imaging survey (Yang et al. 2016 in-prep). In SDSS DR7, a sample of 251 Green Peas were observed as serendipitous spectroscopic targets (Cardamone et al. 2009). A subset of 66 Green Peas have sufficient signal to noise ratio (S/N) in both continuum and emission lines (\Ha, \Hb, and \oiii) to study galactic properties such as SFR, stellar mass, and metallicity (Cardamone et al. 2009; Izotov et al. 2011). Galaxies with an active galaxies nucleus (AGN) (diagnosed by their broad Balmer emission lines or \Ha/[NII] vs. [OIII]/\Hb\ diagram) are excluded.  In Paper I, we matched these 66 Green Peas with the COS archive and studied \lya\ escape in a sample of 12 Green Peas with COS UV spectra. Compared to the larger Green Pea sample, these 12 Green Peas tend to be lower metallicity and lower dust extinction (figure 1). To address the bias and expand the sample size, we took \lya\ spectra of 20 additional Green Peas (PI S. Malhotra, GO 14201). These 20 galaxies were selected based on their metallicity and \ha/\hb\ values to supplement the previous sample, so that the total sample can cover the whole range of metallicity and dust extinction of the parent sample. We use figure 1 to do the selection – first draw grids (shown in figure 1), then pick one or two sources in each grid cell. Note that (a) empty cells are not used and (b) the non-empty cells are not covered perfectly because in the proposal we used gas metallicities measured in Izotov et al. (2011) which are slightly different from the metallicities shown in figure 1. After the selection, we compared the total sample with the parent sample to make sure there is no obvious biases.

We also supplement this sample with 11 additional Green Peas from published literature. In total, we have 43 Green Peas from six HST programs -- 20 galaxies from GO 14201 (PI S. Malhotra), 9 galaxies from GO 12928 (PI A. Henry; Henry et al. 2015), 7 galaxies from GO 11727 and GO 13017  (PI T. Heckman; Heckman et al. 2011; Alexandroff et al. 2015), 2 galaxies from GO 13293 (PI A. Jaskot; Jaskot et al. 2014), and 5 galaxies from GO 13744 (PI T. Thuan; Izotov et al. 2016). 
The 7 galaxies in T. Heckman's program were originally selected as nearby Lyman-break analogs by their high FUV luminosity, high UV flux, and compact size. These 7 galaxies can also be classified as Green Peas by their compact sizes in SDSS images and strong \oiii\ emission lines in SDSS spectra. Their sizes and \oiii\ equivalent width are similar to the Green Peas in Cardamone et al. (2009).
We don't find any obvious bias by including the Lyman-break analogs in the analysis. The 7 Green Peas in A. Jaskot's program and T. Thuan's program were selected as LyC leakers by their extreme [OIII]/[OII] ratios. 
In figure 1, we show the above samples on the metallicity and dust extinction (\ha/\hb\ ratio) diagram. We can see the current sample is a representative Green Pea sample. 

\subsection{Properties from SDSS Spectra}
From SDSS optical spectra of Green Peas, we get many galactic properties. We use the SDSS pipeline measurements of their \ha, \hb, \oiii, and \oii\ emission line fluxes and line width. We correct the measured \Ha\ and \Hb\ fluxes for Milky Way extinction using the attenuation of Schlafly \& Finkbeiner (2011) (obtained from the NASA/IPAC Galactic Dust Reddening and Extinction tool) and the Fitzpatrick (1999) extinction law. Then we calculate E(B-V) assuming the Calzetti et al. (2000) extinction law and an intrinsic \Ha/\Hb\ ratio of 2.86 (if \Ha/\Hb$<2.86$, we set E(B-V)=0), and correct the observed emission line fluxes for dust extinction. We use the stellar mass measured from SDSS spectra by Izotov et al. (2011) for 37 galaxies and the stellar mass in MPA-JHU SDSS catalog for the other 6 galaxies (all are Lyman-break analogs). Note that the methods used in Izotov et al. (2011) and MPA-JHU are different. The masses here should be treated as very rough estimates because it is very hard to get the masses of the underlying old population for these young starburst galaxies.
To measure the metallicity using $T_{e}$ method, we measure the \oiiit\ line flux in SDSS spectra by fitting a Gaussian function to the continuum subtracted \oiiit\ line spectra. 
Then we calculate the metallicity using \oiiit, \oiii, and \oii\ line fluxes following the $T_{e}$ method described in Izotov et al. (2006) and Ly et al. (2014). We convert the extinction corrected \ha\ luminosity to SFR using the formula $SFR(M_{\odot}/yr)=L_{H\alpha}(erg/s)\times10^{-41.27}$ (Kennicutt \& Evans 2012). The dust extinction, mass, metallicity, SFR, and emission lines properties of this sample are shown in Table~1 and Table~2.

\begin{figure}[!ht]
\centering
\includegraphics[width=0.5\textwidth]{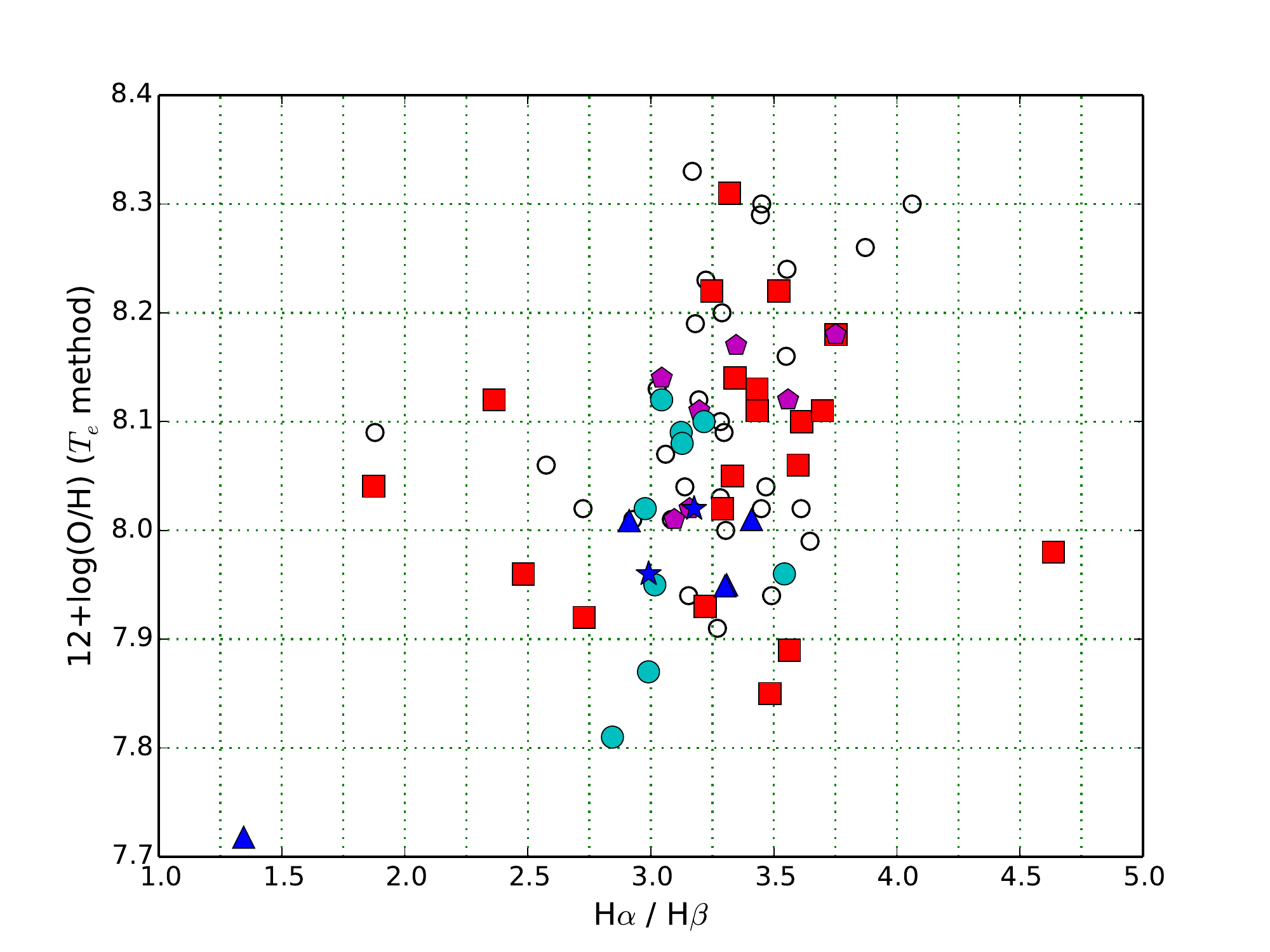}
\caption{The metallicity and dust extinction (\ha/\hb\ ratio) diagram of our Green Pea sample. Red squares shows the 20 galaxies with new HST observations (GO 14201, PI S. Malhotra). The other samples include 9 Green Pea galaxies with low dust extinction (cyan circle, Paper I; Henry et al. 2015), 7 Lyman-break analog galaxies (magenta pentagon, Heckman et al. 2011; Alexandroff et al. 2015), 2 Lyman-continuum leaker candidates (blue star, Jaskot et al. 2014), and 5 confirmed Lyman-continuum leakers (blue triangle, two blue triangles overlap; Izotov et al. 2016). The black hollow circles show the other galaxies without HST UV spectra in the sample of 66 Green Peas. Note that a few sources have very small Ha/Hb values. The reasons are not yet well understood, but could be 1) poor flat-field calibration or sky subtraction, 2) different gas conditions from the case-B assumption.}
\end{figure}

\subsection{HST/COS Observation}
In our program GO14201, we used HST/COS to observe 20 Green Peas with one orbit per target. First, the targets were imaged in the COS acquisition mode ACQ/IMAGE with MIRRORA, from which we got high resolution near-UV (NUV) images. The targets were centered accurately (error $\sim$ 0.05\arcsec) in the 2.5\arcsec\ diameter Primary Science Aperture. Then the spectra were taken with grating G160M to cover rest-frame wavelength ranges about $1100-1400$ \AA. 
The other archival Green Peas in our sample were also observed in the same COS acquisition mode ACQ/IMAGE with MIRRORA, and their spectra were taken with grating G130M and/or G160M.  
The NUV acquisition images of this sample are shown in figure 2. 

The spectral resolution of the above observation is about FWHM$\sim$20 \kms\ for a point source (James et al. 2014). The actual spectral resolution depends on source angular sizes. The half-light radius of the NUV emission of Green Peas are about 10 pixels (dispersion $\sim$ 0.012 \AA\ $pixel^{-1}$) and it results in FWHM$\sim$40 \kms\ for the UV continuum spectra. As the \lya\ sizes of Green Peas are somewhat larger than the UV continuum sizes (Yang et al. 2016b), the spectral resolutions are worse for the \lya\ emission lines. We retrieved COS spectra of this sample from the HST MAST archive after they were processed through the standard COS pipeline.

\begin{figure*}[!ht]
\centering
\includegraphics[width=0.9\textwidth]{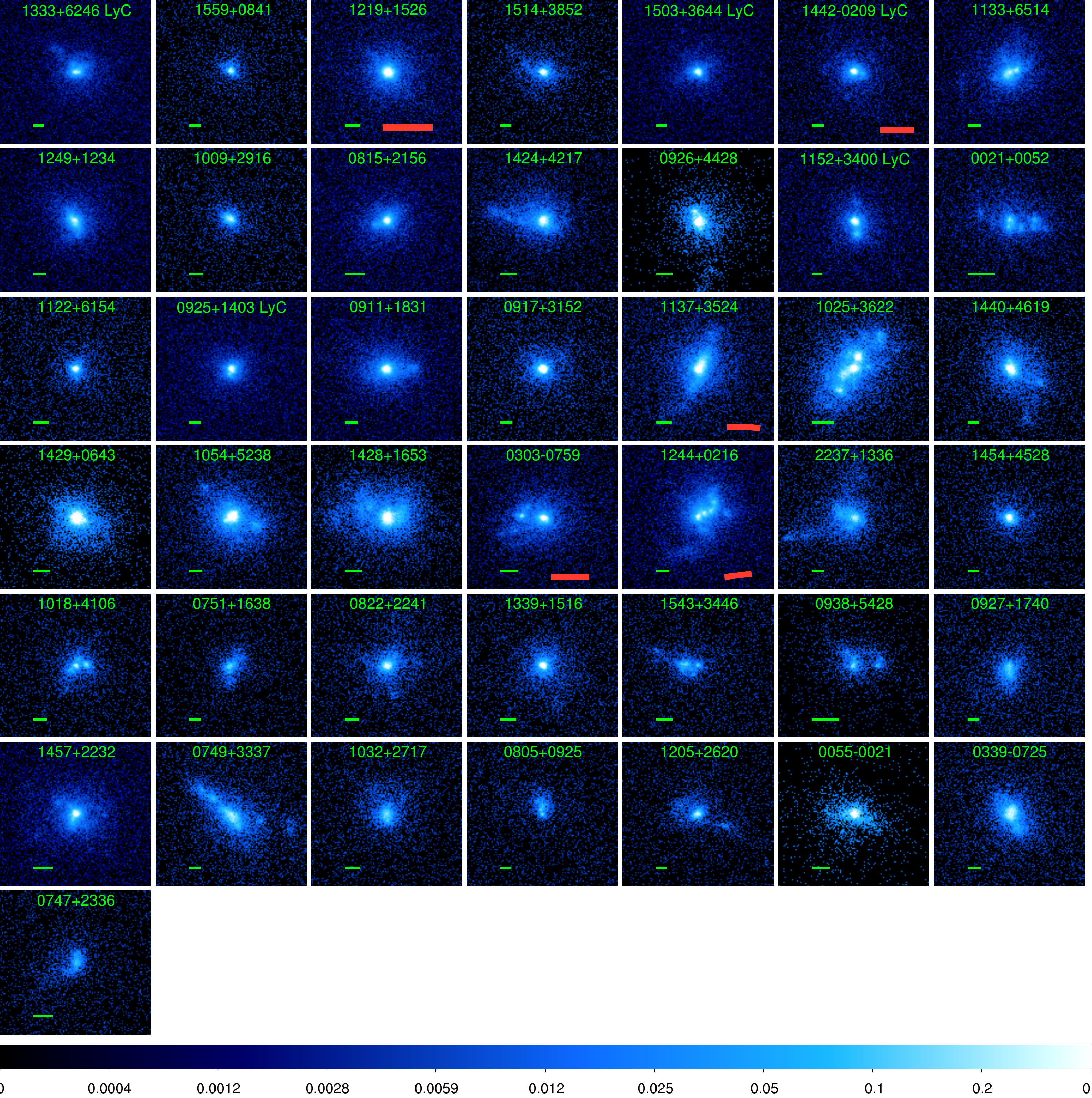}
\caption{The $3\arcsec\times3\arcsec$ NUV images of Green Peas from the COS target acquisitions. In all panels, the colors are in log-scale with the same count-rates limits (from 0 to 0.4). These images are sorted by decreasing \fesc\ from left to right, and from top to bottom. The label shows the ID of each Green Pea. The five LyC leakers are marked with `LyC'. The green bar in each panel shows the physical scale of 1 Kpc.}
\end{figure*}

\section{\lya\ Equivalent Width and Escape Fraction}
\subsection{Measurements of \lya\ flux, EW, and escape fraction}
Most Green Peas in our sample show strong \lya\ emission lines (figure 3). But about 1/3 Green Peas have relatively weak \lya\ lines, where the \lya\ absorptions in underlying continuum become non-negligible. Since we want to measure \lya\ emission from the recombination of interstellar HI gas, we need to subtract the underlying continuum. 

We first estimate a constant local continuum from wavelength ranges near \lya\ where the spectra look flat and there are no obvious emission or absorption features. We calculate the ``local continuum" $f_{\lambda}$(continuum) as the average of the spectra in these continuum ranges. 

For 33 Green Peas without damped \lya\ absorption (see Table 3), we subtract the ``local continuum" and calculate the \lya\ flux by integrating the spectra in wavelength range $\sim~1212-1221$ \AA. Then we correct the \lya\ flux for underlying stellar absorption. The equivalent width of stellar \lya\ absorption mostly depends on the star formation history and age of the stellar population (Pena-Guerrero \& Leitherer 2013). By comparing the \ha\ EW of these Green Peas (about $300-900$\AA) with model predictions of \ha\ EW in star-forming galaxies, we found that these Green Peas probably have instantaneous starburst with a burst age of $4-5$ Myr (Levesque \& Leitherer 2013). According to the model calculations in Pena-Guerrero \& Leitherer (2013), the stellar \lya\ absorption EW is about $-7$ \AA. So we correct the \lya\ fluxes of these 33 Green Peas by an EW=$-7$ \AA\ absorption.  

In another 8 Green Peas, the spectra show damped \lya\ absorption wings and weak residual \lya\ emission lines. The damped \lya\ absorption is caused by {\it interstellar absorption of the continuum and/or the \lya\ absorption of the underlying stellar atmosphere continuum spectra}. 
To measure flux of the residual \lya\ emission, we subtract \lya\ line spectra by a constant ``absorbed continuum". The ``absorbed continuum" is estimated as the average in the wavelength range where the \lya\ emission line meets the absorbed continuum. Then we integrate the \lya\ line spectra to get \lya\ flux. Since the above absorption correction already includes stellar \lya\ absorption, we don't need to correct the stellar absorption for these 8 Green Peas. Note that in some cases, the stellar absorption might have a very narrow component which is not fully corrected by this method.

In the remaining two Green Peas (GP0339$-$0725 and GP0747$+$2336), the \lya\ lines are too weak and we didn't detect \lya\ emission.  

Then we correct the measured \lya\ fluxes for Milky Way extinction using the Fitzpatrick (1999) extinction law. 
The rest-frame EW(\lya) is calculated using the \lya\ fluxes and the ``local continuum" as EW(\lya)=flux(\lya)/$f_{\lambda}$(continuum)/(1+redshift). 
The \lya\ escape fraction, \fesc, is defined as the ratio of the measured \lya\ flux to the intrinsic \lya\ flux. Assuming case-B recombination, the intrinsic \lya\ flux is about 8.7 times dust extinction corrected \Ha\ flux (See Henry et al. 2015 for discussions about the factor 8.7). Thus the \fesc\ is \lya(observed)/(8.7$\times H\alpha_{corrected}$).  The SDSS \Ha\ spectra were taken with 3\arcsec\ diameter aperture which matches the COS 2.5\arcsec\ diameter aperture very well. Note that many \lya\ galaxies have a very extended \lya\ halo (e.g. Ostlin et al. 2009; Hayes et al. 2013; Momose et al. 2014). For these Green Pea galaxies, their \lya\ to UV size ratios are about 2$-$4 (Yang et al. 2017). Thus COS 2.5\arcsec\ aperture probably captured the majority of \lya\ emission of those Green Peas.

Because the total counts per pixel in the UV continuum of this sample are small, we calculate the error spectra using the Poisson noise of the total counts. The statistical errors of \lya\ fluxes are calculated from the error spectra using the error propagation formula. The \lya\ flux, luminosity, EW(\lya), and \fesc\ are  shown in Table 3. A comparison of the \fesc\ and EW(\lya) is shown in figure 4.

\begin{figure*}[!ht]
\centering
\includegraphics[width=\textwidth]{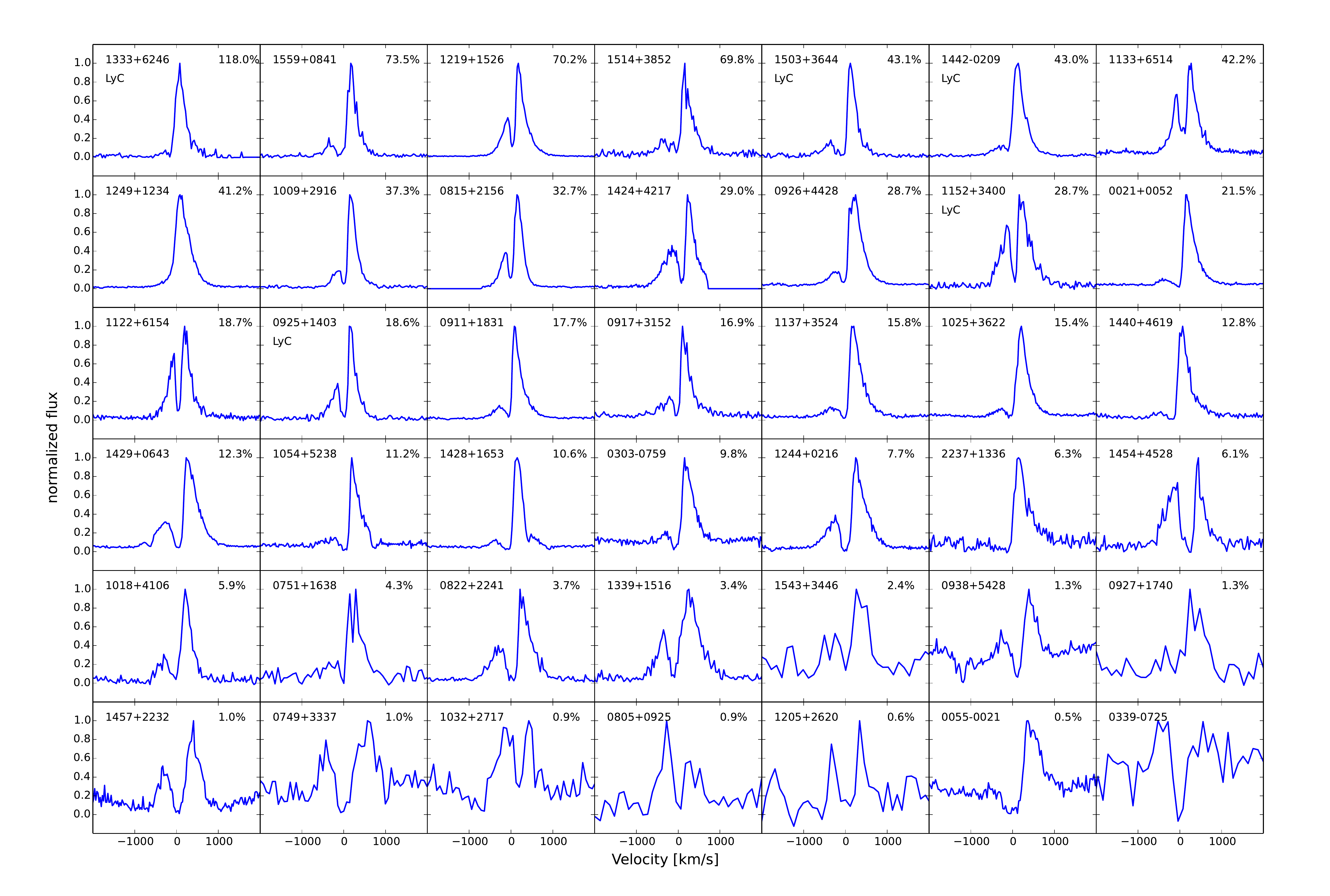}
\caption{\lya\ emission line spectra of Green Peas before subtracting continuum. These 42 galaxies are sorted by decreasing \fesc\ from left to right, and from top to bottom. The ID and \fesc\ are given in each panel. The five LyC leakers are marked with `LyC'. The last one galaxy (GP0339$-$0725) shows weak \lya\ absorption. One Green Pea (GP0747$+$2336) is not shown here, because its \lya\ spectra is very noisy and no \lya\ emission or absorption lines are detected.}
\end{figure*}

\begin{figure}[ht]
\centering
  \includegraphics[width=0.45\textwidth]{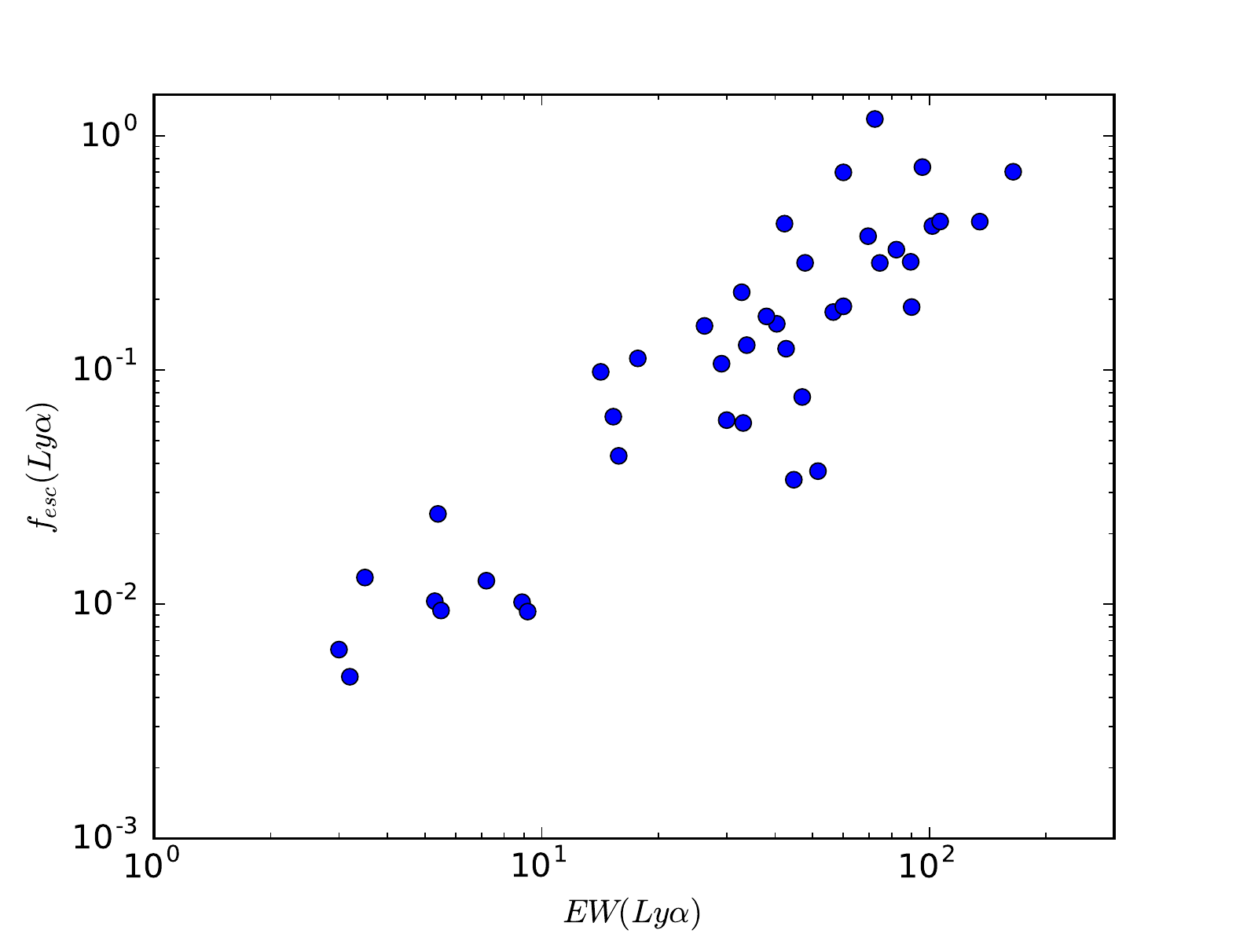}
 \caption{Comparison of the \fesc\ and EW(\lya) of Green Peas. }
\end{figure}

\subsection{\lya\ EW distribution of Green Peas}
With a large sample of Green Peas that cover the whole ranges of dust and metallicity, we now have a more reliable estimation of the EW(\lya) distribution of Green Peas than previous result. 
41 out of 43 Green Peas show \lya\ emission lines. 28 out of 43 GPs (65\%) in our sample have rest-frame EW(\lya) $\gtrsim$ 20\AA\ and would be classified as LAEs in a typical high-redshift narrow-band survey. We compared the EW(\lya) distribution of these 28 Green Peas to high redshift LAEs samples. The high redshift LAEs samples include a sample of $z=2.8$ narrow-band selected LAEs (Zheng et al. 2016) and a sample of spectroscopically confirmed LAEs at $z$=5.7 or 6.5 (Kashikawa et al. 2011). To be consistent with the methods used in high-$z$ LAEs studies, we use the EW(\lya) of Green Peas without correction of the stellar \lya\ absorption. We also add a GALEX selected $z\sim0.3$ LAE sample to the comparison (Deharveng et al. 2008; Cowie et al. 2011; Finkelstein et al. 2009; Scarlata et al. 2009). Figure 5 shows the cumulative EW(\lya) fraction distributions of these four samples. 
These 28 Green Peas have very similar EW(\lya) distribution to the high-redshift ($z=2.8$) sample. So Green Peas in general are the best nearby analogs of high-$z$ LAEs.

\begin{figure}[ht]
\centering
  \includegraphics[width=0.45\textwidth]{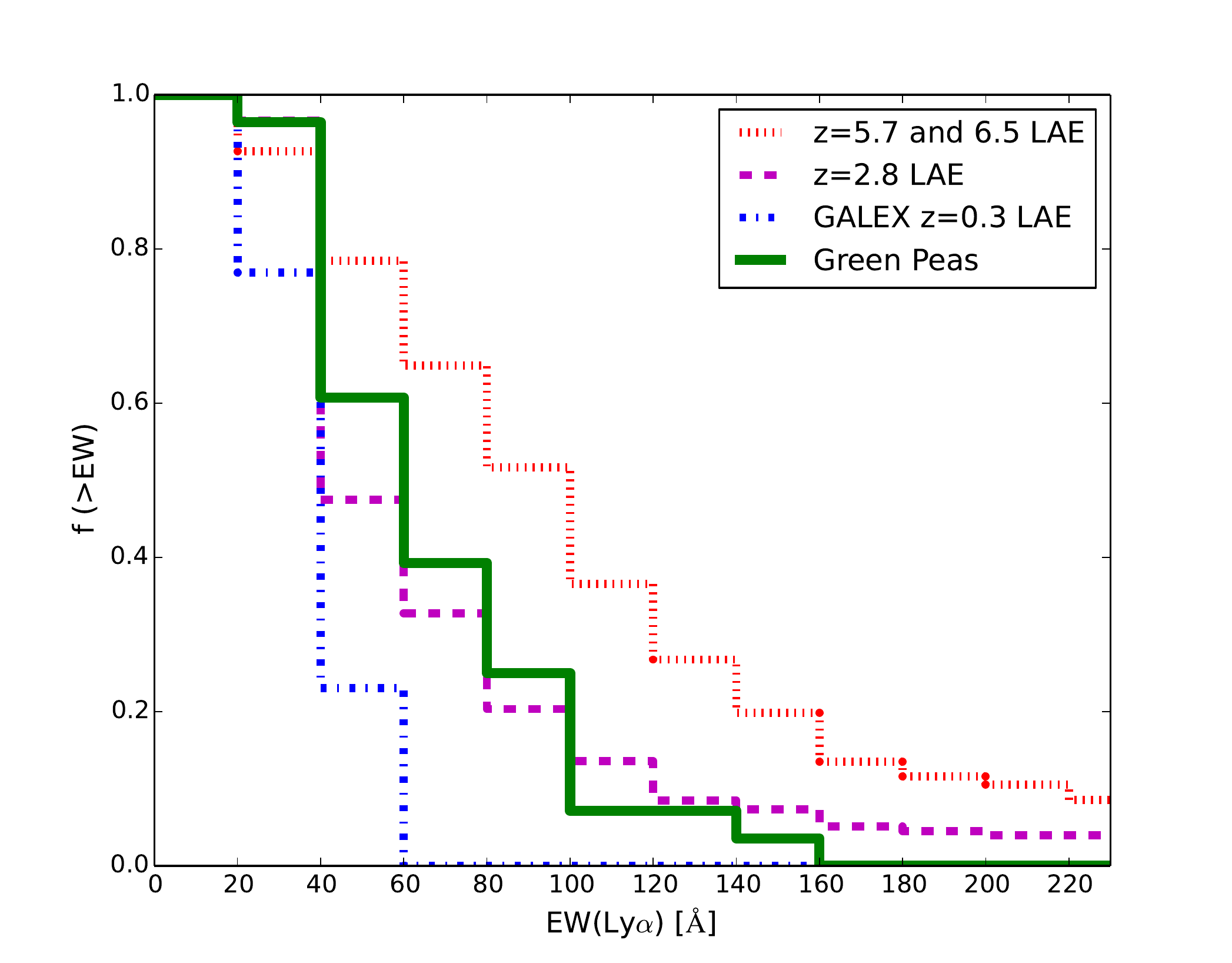}
 \caption{Here we compare the rest-frame EW(\lya) distribution of Green Peas with different samples. The solid green line shows the sample of 28 Green Peas with EW(\lya) $\gtrsim$ 20\AA. The blue dash-dot line shows the GALEX $z\sim0.3$ LAE sample (Cowie et al. 2011; Finkelstein et al.  2009; Scarlata et al. 2009).  The magenta dashed line shows the $z=2.8$ LAE sample from Zheng et al. (2016). The red dotted line shows the $z=5.7$ and $6.5$ LAE sample from Kashikawa et al. (2011).}
\end{figure}

\section{\lya\ escape and \lya\ profiles}
\subsection{Kinematic Features of \lya\ Profile}
In the \lya\ escape process, \lya\ photons are resonant scattered by the HI gas. Depending on the column density and bulk motion of HI gas, the resonant scatterings can significantly modify the \lya\ profile. Therefore the Lya profile carries a lot of information about the HI gas properties. High-$z$ LAEs usually show an asymmetric or a double-peaked \lya\ emission line profile (e.g. Rhoads et al 2003; Kashikawa et al. 2011; Erb et al. 2014). For LAEs with detected optical emission lines and systemic redshifts, the peaks of \lya\ profiles are usually redshifted with respect to the systemic velocities (McLinden et al. 2011, 2014; Chonis et al. 2013; Hashimoto et al. 2013; Song et al. 2014; Shibuya et al. 2014; Erb et al. 2014). The velocity offset of \lya\ emission line from the systemic velocity is usually smaller in LAEs than in continuum selected galaxies with weaker \lya\ emission lines or \lya\ absorptions (Shapley et al. 2003). 

Most Green Peas show double-peaked \lya\ profiles (figure 3). For a typical double-peaked profile, we define the ``red peak'' as the peak in the \lya\ line profile occurring at velocity \textgreater\ 0,  the ``blue peak'' as the \lya\ peak at velocity \textless\ 0,  and the ``valley'' as the flux minimum between the two peaks.

With a sample covering a large range of properties, we can see the \lya\ profiles are diverse. 
In figure 3, the 42 Green Peas are sorted by decreasing \fesc\ from top left to bottom right. Three Green Peas with high \fesc\ show single peak profiles where the peak velocities are close to zero (GP1333$+$6246, GP1442$-$0209, and GP1249$+$1234). Many Green Peas with intermediate \fesc\ generally show double-peaked profiles with much stronger red peaks than blue peaks. On the other hand, many Green Peas with low \fesc\ have a relatively large ratio of blue peak to red peak. 

As in Paper I, we measure four kinematic features of the \lya\ profile: i) the blue peak velocity V(blue-peak); ii) the red peak velocity V(red-peak); iii) the peak separation V(red-peak)$-$V(blue-peak); and iv) the full width at half maximum (FWHM) of the red portion of \lya\ profile, FWHM(red). The velocities are relative to the systemic redshift derived from SDSS spectra. 
The measurements of these kinematic features are shown in Table~3. 
For some Green Peas, we don't measure their velocities because their \lya\ profiles are too noisy. In the notes of Table 3, we explain the reason for each profile without velocity measurement. 
To measure the errors of velocity peaks, we use a Monte-Carlo method to generate 1000 fake spectra by adding Gaussian noise (with the error spectra as the $\sigma$ of Gaussian noise) to the observed spectra. Then we measure the peak velocities of these 1000 fake spectra and use the standard deviations as the errors. 
In summary, we have measurements of V(blue-peak) and the peak separation in 28 galaxies, and of V(red-peak) and FWHM(red) in 37 galaxies.

\begin{figure}[!ht]
\centering
\includegraphics[width=0.5\textwidth]{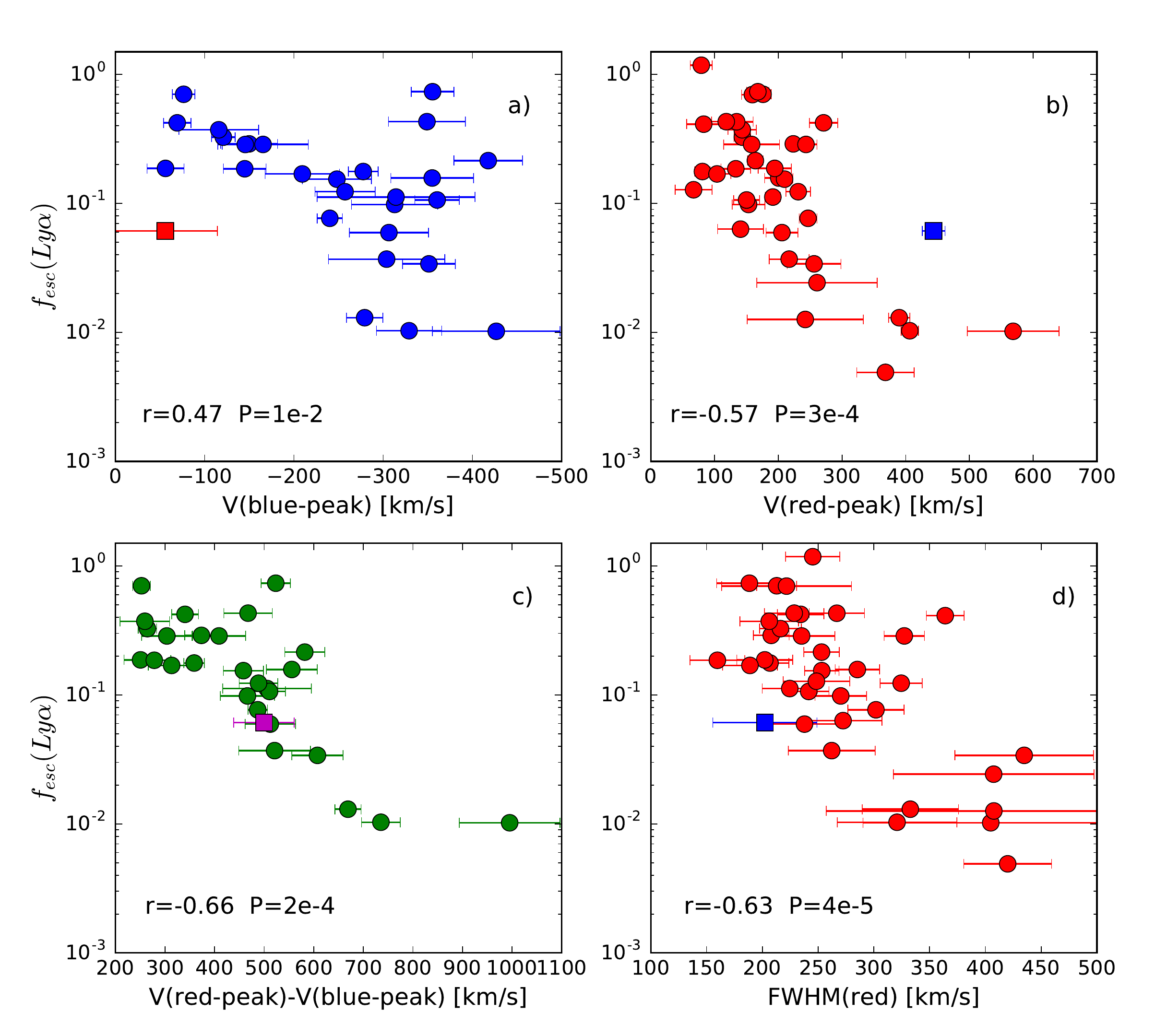}
\caption{Relations between \fesc\ and the kinematic features of \lya\ profile: (a) \fesc\ and the blue peak velocity of \lya\ profile, V(blue-peak);  (b) \fesc\ and the red peak velocity of \lya\ profile, V(red-peak); (c) \fesc\ and the peak separation of \lya\ profile; (d) \fesc\ and the FWHM of the red portion of \lya\ profile, FWHM(red). The Spearman correlation coefficient and null probability are shown. GP1454+4528 with possible gas inflows is marked by a square in different color in each panel.}
\end{figure}

\subsection{Relations between \lya\ escape and \lya\ kinematics}
We show the relations between \fesc\ and the kinematic features of \lya\ profiles in figure 6. As \fesc\ covers a range  of about 3 dex, we show it in logarithmic scale. \fesc\ shows anti-correlations with all four kinematic features -- V(blue-peak), V(red-peak), the peak separation V(red-peak)$-$V(blue-peak), and the FWHM(red). We calculate the Spearman correlation coefficients of these relations (shown in each panel of figure 6). 

In Paper I, we found the \fesc\ correlates strongly with V(blue-peak). Here we can see most Green Peas still follow the correlation, but there are a few Green Peas with large scatter. So the overall correlation is worse than in Paper I. These outliers suggest that the \lya\ blue peak velocities are determined by multiple mechanisms.
For example, one outlier (GP1454+4528, marked with a square and different color in figure~6) has a distinct profile with the largest positive V(valley) (the velocity at the inter-peaks dip) and very strong blue portion \lya\ emission. Its V(blue-peak) and V(red-peak) clearly offset from the trends. However, if we exchange the V(blue-peak) and V(red-peak), then it follows the trends very well.
There is probably strong gas inflows as well as gas outflows in this galaxy.
We excluded this object from the calculation of correlation coefficients. 

On the other hand, in Paper I, we found large scatter between \fesc\ and V(red-peak) with 12 Green Peas. However, as the current sample covers a large range of \fesc\ and V(red-peak), \fesc\ shows an anti-correlation with V(red-peak). 
The relation between \fesc\ and V(red-peak) in this Green Peas sample is very similar to the relations between EW(\lya) and V(red-peak) in high redshift LAEs and LBGs, where the LAEs have high EW(\lya) and small V(red-peak), while the LBGs have small EW(\lya) and large V(red-peak) (Shapley et al. 2003; Hashimoto et al. 2013; Erb et al. 2014). 

We also found that \fesc\ anti-correlates with FWHM(red). We do a linear fit to this relation and get the following function.
\vspace{0.1cm}
\[ log(f^{Ly\alpha}_{esc})=-0.545\times(FWHM(red)/100 km/s) + 0.563\] 
\vspace{0.1cm}
The scatter of this relation is 0.43 dex in log(\fesc). Since any high-$z$ LAE with a spectrum will have a measured FWHM for the red peak, it is easy to use this relation to infer the \lya\ escape fraction of high-$z$ LAE.

{\em Brief interpretations:} The \lya\ profile depends on the column density and the kinematics of HI gas. As the HI column density increases, the numbers of scatterings for \lya\ photons increase. The more scatterings generally result in larger offsets of peak velocities (V(blue-peak) and V(red-peak)) and broader line profile (FWHM(red)). Also, more scatterings increase the \lya\ photons' path lengths which makes the \lya\ radiation more susceptible to dust extinction and consequently decreases the \lya\ escape fraction. Thus those anti-correlations mostly indicate that the \fesc\ decreases as the column density of HI gas increases.

\begin{figure}[ht]
\centering
  \includegraphics[width=0.5\textwidth]{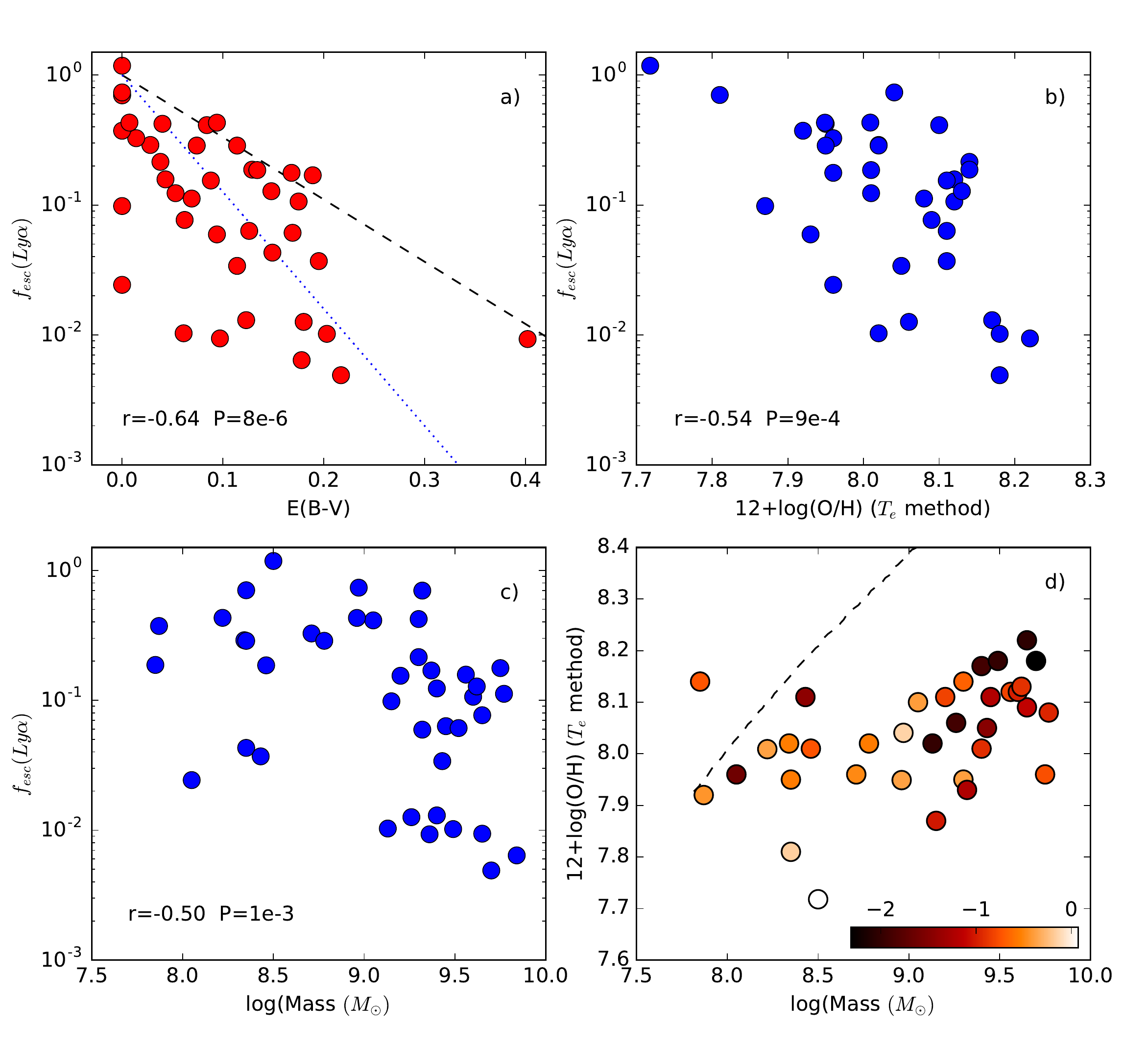}
 \caption{a) Relation between \fesc\ and dust extinction E(B-V). The black dashed (blue dotted) line shows the expected \lya\ escape fraction if \lya\ is only absorbed by dust following the Calzetti et al. (2000) extinction law (the SMC extinction law). b) Relation between \fesc\ and the metallicity from $T_{e}$ method. c) Relation between \fesc\ and stellar mass. The Spearman correlation coefficient and null probability are shown in panel a), b), and c).  d) The mass-metallicity relation of this sample. The color-bar shows the value of log(\fesc). The dashed line shows the mass-metallicity relation for SDSS galaxies in Amorin et al. (2010).}
\end{figure}

\section{\lya\ escape and other galactic properties}

\subsection{dust extinction, stellar mass, and metallicity}
These Green Peas are very well studied galaxies and provide a great opportunity to explore the dependence of \lya\ escape on other galactic properties. 
Previous studies have found that \fesc\ anti-correlates with dust extinction (Atek et al. 2014; Cowie et al. 2011; Paper I). However the relation between \fesc\ and metallicity are unclear (Finkelstein et al. 2011; Atek et al. 2014; Hayes et al. 2014; Paper I). 
Our sample covers the full ranges of dust extinction and metallicity of Green Peas. In figure~7, we show the relations between \fesc\ and E(B-V), metallicity, and stellar mass. The Spearman correlation coefficients of these relations are shown figure~7.

The Green Peas with higher dust extinction tend to have smaller \fesc, confirming that dust extinction is an important factor in \lya\ escape. In figure 7a, we also show the expected \lya\ escape fractions if \lya\ is only absorbed by dust following the Calzetti et al. (2000) extinction law (dashed line) or the SMC extinction law (Gordon et al. 2003) (dotted line). The SMC extinction law is steeper in FUV  than the Calzetti et al. (2000) extinction law, so the extinction of \lya\ emission is larger for SMC extinction law. 
Many Green Peas are below the dashed and dotted lines, because resonant scatterings increase the escape path length of \lya\ photons and the chances of being absorbed by dust.
Interestingly, many Green Peas are above the relation for SMC extinction law. If the dust extinction in Green Peas follows SMC extinction law, then it probably suggests resonant scatterings in clumpy dust distributions {\it decrease} the dust extinction of \lya\ emission (Neufeld 1991; Hansen \& Oh 2006; Finkelstein et al. 2009; Scarlata et al. 2009; but also see Laursen et al. 2013 showing that clumpy media does not decrease the dust extinction of \lya\ for typical conditions in LAEs).

\fesc\ also anti-correlates with metallicity and stellar mass. In the \fesc\ vs. metallicity diagram, only 37 galaxies with \oiiit\ line $S/N>3$ are shown. In figure~7, we also show the mass-metallicity relation of Green Peas and color the sample with \fesc. The dashed line shows the mass$-$metallicity relation for SDSS galaxies in Amorin et al. (2010), where the metallicity of SDSS galaxies are calculated with the same effective temperature method. These Green Peas have lower metallicities than the mass$-$metallicity relation of SDSS galaxies, similar to other emission line selected galaxies (Xia et al. 2012; Ly et al. 2014; Song et al. 2014).  These Green Peas with lower metallicities and smaller masses have less dust extinction. In addition, ionized gas outflows can blow out the metal enriched gas and decrease the metallicity and dust extinction. At the same time, the ionized gas outflows can make holes with low HI column densities and help \lya\ escape.

\subsection{Morphology and size of UV emission}
We get the NUV image of each object from the COS target acquisition (figure 2). So we also explore the relation between \lya\ escape and the UV morphology. 
The pixel scale of NUV image is 0.0235 $\pm$ 0.0001 arcsec/pixel. The FWHM of point spread function is about 2 pixels or 0.047\arcsec.  
As we can see from the images, most Green Peas are very small and compact. Multiple clumps, tidal tails, and asymmetric shapes are common, which may suggest {\it dwarf-dwarf mergers} are common in Green Peas. In figure 2, these images are sorted by decreasing \fesc\ from left to right, and from top to bottom. The \fesc\ does not show an obvious relation with the morphology. 

We then use GALFIT (Peng et al. 2010) to measure the galaxy size. We fit the image with a single Sersic profile component and get the half light radius of each galaxy. The half light radii are shown in Table~1. The relation between \fesc\ and the half light radius has very large scatter.

\begin{figure}[ht]
\centering
  \includegraphics[width=0.5\textwidth]{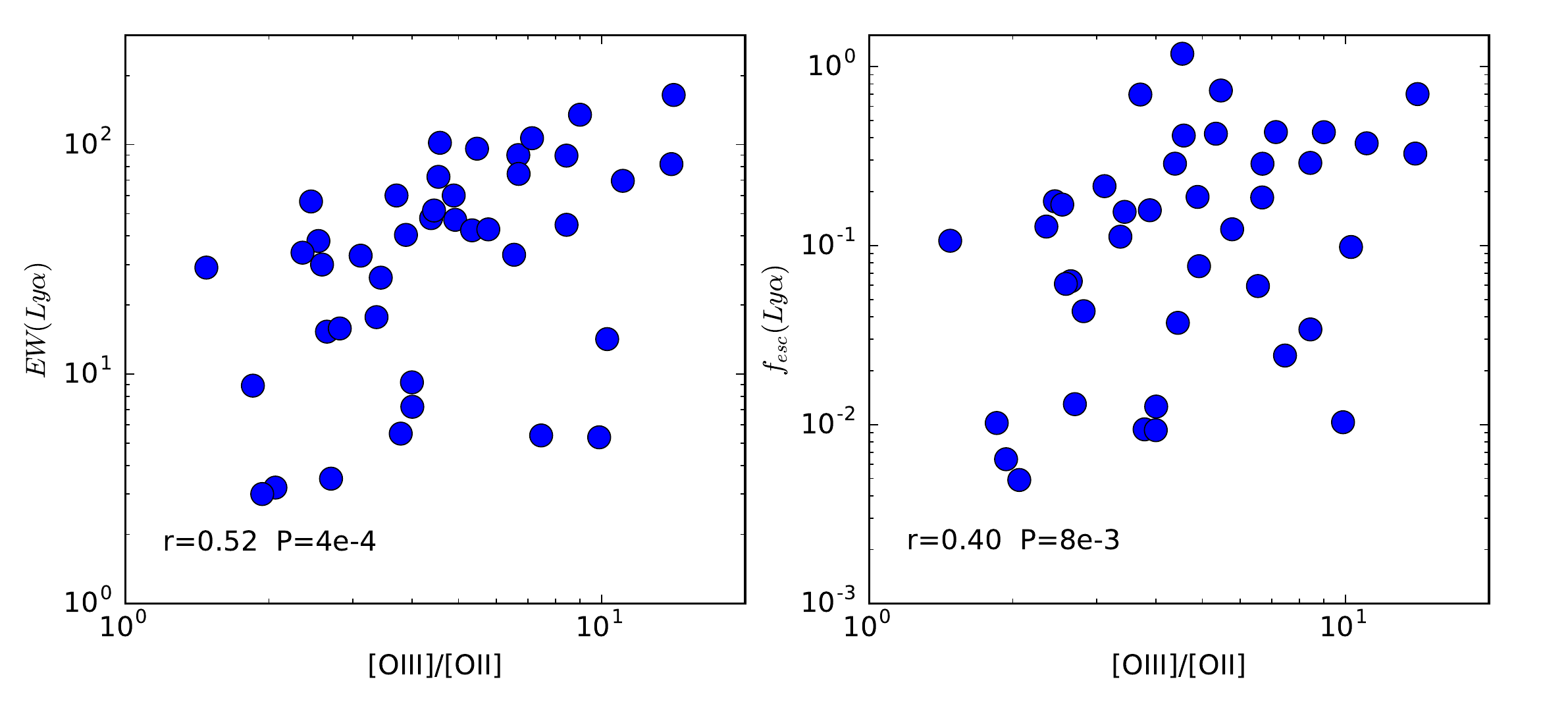}
 \caption{Left: Relation between EW(\lya) and [OIII]/[OII]. Right: Relation between \fesc\ and [OIII]/[OII].  [OIII]/[OII] is defined as ([OIII]$\lambda$4959+[OIII]$\lambda$5007)/([OII]$\lambda$3726+[OII]$\lambda$3729). The Spearman correlation coefficient and null probability are shown in each panel.}
\end{figure}

\begin{figure*}[ht]
\centering
  \includegraphics[width=1.0\textwidth]{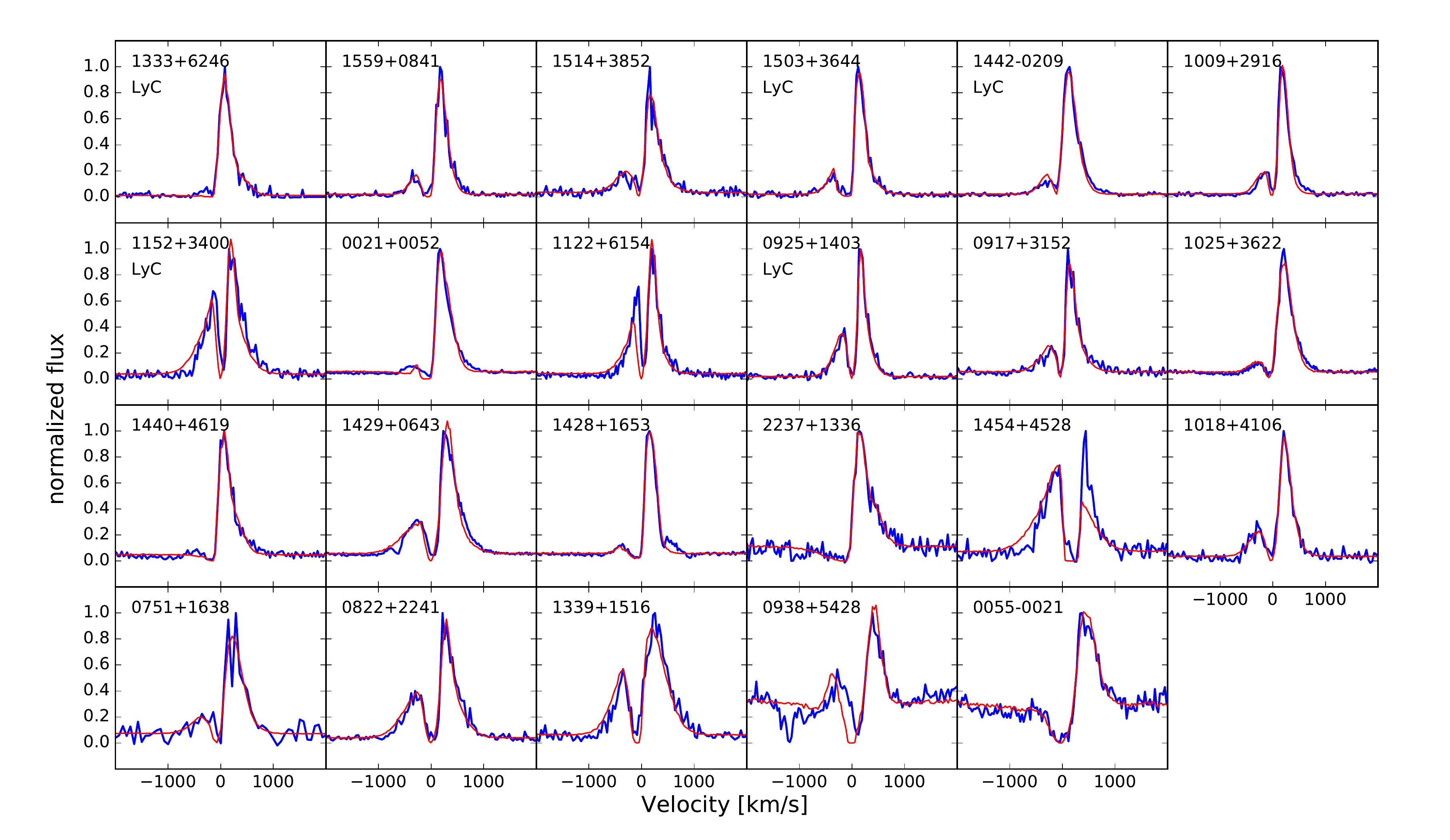}
 \caption{The observed \LyA\ profiles (blue lines) and the best fit \LyA\ profiles (red lines) for 23 Green Peas with good S/N in their \lya\ profiles. In Paper I, we showed the radiative transfer model fitting results of another 12 Green Peas. These galaxies are sorted by decreasing \fesc\ from left to right, and from top to bottom.}
\end{figure*}

\subsection{[OIII]/[OII] ratio}
Green Peas are selected to have large [OIII]/[OII] ratios. The [OIII]/[OII] ratio has been used to select LyC leaker candidates, and large  [OIII]/[OII] may indicate the existence of paths with low HI optical depth (Jaskot \& Oey 2014; Izotov et al. 2016). In figure 8, we show the relations of EW(\lya) vs. [OIII]/[OII] and \fesc\ vs. [OIII]/[OII]. The \lya\ line strength generally increases with [OIII]/[OII], but the scatter is large.

\section{\lya\ profile fitting}
The \lya\ emission line profiles can usually be explained by resonant scatterings of \lya\ photons by an outflowing HI gas shell (e.g. Ahn et al. 2001; Verhamme et al. 2006; Dijkstra et al. 2006; Schaerer et al. 2011). To extract more information from the \LyA\ profiles and explore the physical process of \LyA\ escape, we fit the \LyA\ profiles with the outflowing HI shell radiative transfer model (Dijkstra et al. 2014; Gronke et al. 2015).

In the model, \LyA\ photons were generated by a source fully surrounded by a spherical dusty HI gas shell which scattered/absorbed the \lya\ photons. The intrinsic \LyA\ line has a Gaussian profile with width $\sigma$. The shell is described by four parameters:  (i) outflow velocity $v_{exp}$, (ii) HI column density $N_{HI}$, (iii) temperature T (including turbulent motion as well as the true temperature), and (iv) dust optical depth $\tau_{d}$. Generally, these parameters affect the \lya\ profile as follows: a larger outflow velocity and a smaller $N_{HI}$ will decrease the red-peak velocity; a higher temperature will generally broaden the line profile; a larger dust optical depth will decrease the line strength. 
Then we find the best-fit model parameters ($\sigma$, $v_{exp}$, $N_{HI}$, T, $\tau_{d}$) and calculate the errors of parameters with Markov Chain Monte Carlo (MCMC) method. We refer the reader to Gronke et al. (2015) and Paper I for details of the model and the fitting method. 

In Paper I, we showed the fitting results of 12 Green Peas. The model fit nine profiles very well, but failed in the other three profiles. Here we show the fitting results for another 23 Green Peas (out of the 31 additional Green Peas) with sufficient S/N in their \lya\ profiles. The model fit the observed profiles very well in many cases (figure 9). The best fit parameters are shown in Table 4. We discussed a few interesting fitting results below. 

{\em (1) HI column density:} 
In Paper I, we found \fesc\ anti-correlates with the best fit $N_{HI}$ for the 12 Green Peas. Here we show the relation between \fesc\ and the best fit $N_{HI}$ in figure~10 for the combined sample of 35 Green Peas. The result confirms the  anti-correlation between \fesc\ and $N_{HI}$. For the three cases (GP1424+4217, GP1133+6514, and GP1219+1526, marked by large blue circles) where the fitting procedure failed, we plot the $N_{HI}$ obtained by manually adjusting the model parameters to match the observed depth of the ``valley" and the relative heights of blue and red peaks (see Section 6 of Paper I). For GP1454$+$4528 (marked by a red square) with gas inflow, the fitting was bad. For the two galaxies marked by large cyan triangles, the best fit $N_{HI}$ are not constrained. If the three galaxies marked by the square and triangle are excluded, the Spearman correlation coefficient for the relation of \fesc\ and $N_{HI}$ is r=-0.59 (P=4e-4).  If all six galaxies marked by the large circle, square and triangle are excluded, the Spearman correlation coefficient  is r=-0.52 (P=4e-3).   
This result is consistent with studies of high redshift LAEs that suggested LAEs have lower $N_{HI}$ than non-LAEs (e.g. Shibuya et al. 2014; Erb et al. 2014; Hashimoto et al. 2015).  
Therefore the low column density of HI gas is a key factor to make \LyA\ escape.

\begin{figure}[ht]
\centering
  \includegraphics[width=0.5\textwidth]{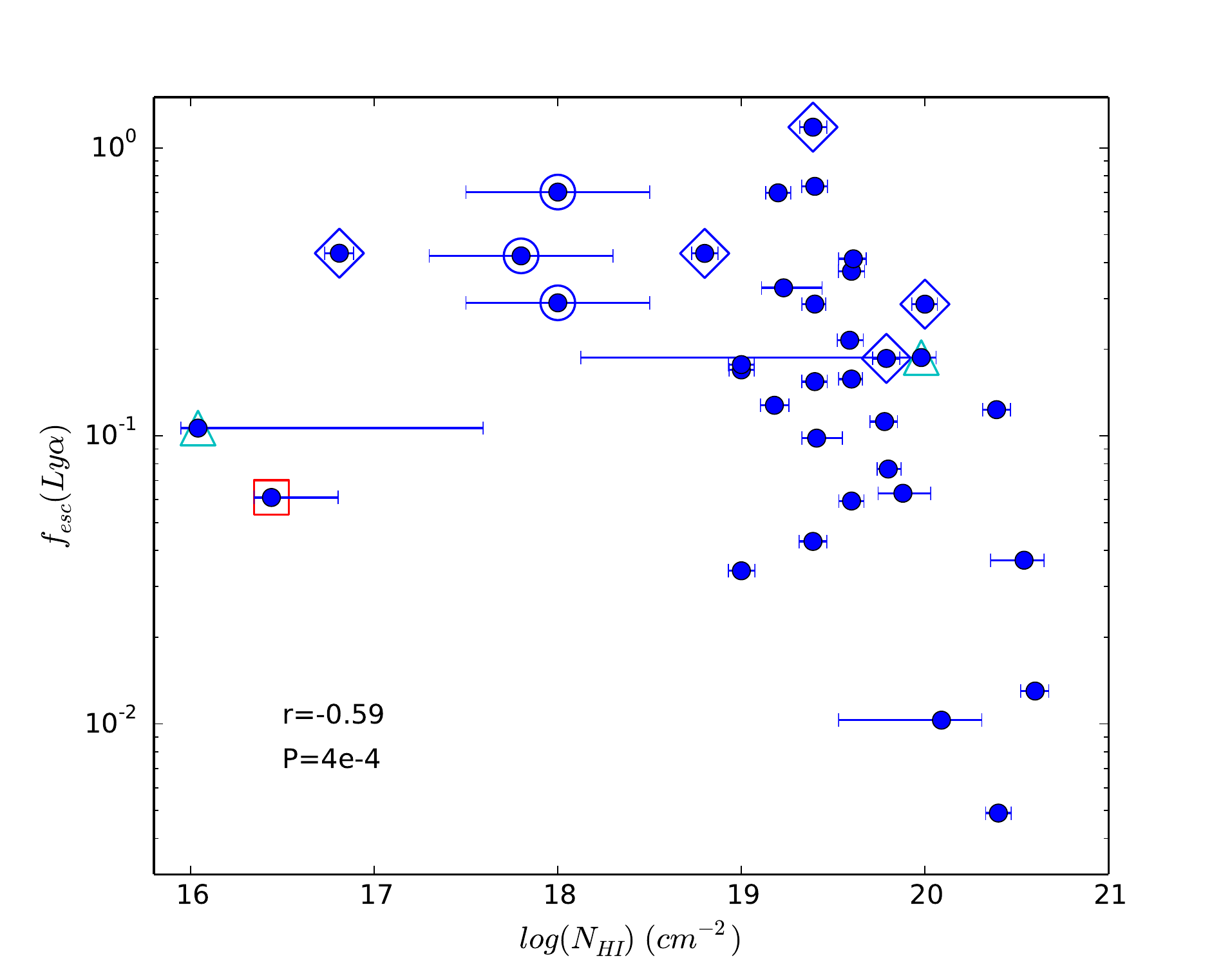}
 \caption{Relation between \fesc\ and the best fit $N_{HI}$ from radiative transfer model. Five known LyC leakers are marked by large diamonds. For the three cases (GP1424+4217, GP1133+6514, and GP1219+1526, marked by large blue circles) where the fitting procedure failed, we plot the $N_{HI}$ obtained by manually adjusting the model parameters to match the observed depth of the ``valley" and the relative heights of blue and red peaks (see Section 6 of Paper I). For GP1454$+$4528 (marked by a large red square) with gas inflow, the fitting is bad (see figure~8). For the two galaxies marked by large cyan triangles (GP1428$+$1653 and GP1122$+$6154), the best fit $N_{HI}$ are not constrained. The Spearman correlation coefficient is calculated without the three galaxies marked by square and triangle.}
\end{figure}

{\em (2) Intrinsic \lya\ line width:} 
The intrinsic \lya\ line Gaussian width $\sigma$ is about $2-3$ times larger than the \ha\ Gaussian width in many cases, as we discussed in Paper I.  In four cases, the best fit $\sigma$ is narrow and comparable to the \ha\ width because the best fit profile only has a single peak. 
The wide intrinsic \lya\ line profile can be due to important radiative transfer effects that broaden \lya\ profile near to the source, before the processes attributed to the outflowing HI shell. 

{\em (3) Outflow velocities:} 
The best fit shell outflow velocities are mostly between 5 to 170 \kms\ which are generally smaller than the outflow velocities measured from the low-ionized UV absorption lines (Yang et al. in-prep). 
This may suggest the low-ionized absorption lines trace a different gas component from the HI gas. 
We also noticed that for six profiles with strong blue peaks, the best fit shell outflow velocities are smaller than 20 \kms. In GP1454$+$4528, the outlier discussed in section 4.1, the HI gas shell is inflowing with a best-fit velocity of $171$ \kms. 

{\em (4) The three failed cases:} In Paper I, the model failed in three profiles with positive velocities at the line ``valley". We later improved the model by adding a shift of the velocity zero point as a free parameter of the fitting. The improved model can fit these three profiles very well. But the shifts of velocity zero points are about $90 - 150$ \kms\ which are too large to be due to the errors of wavelength calibration. Those large shifts may be explained by some additional radiative transfer effects before the \lya\ photons meet the HI gas shell.

Although the shell model captures many real radiative transfer effects and can fit the \lya\ profiles very well, we should be cautious about the interpretation of the best fit parameters. A simple shell model can mimic more complex real physical properties (Gronke et al. 2016). For example, a low $N_{HI}$ model can mimic a model in which the gas is clumpy and the covering factor is low (Gronke \& Dijkstra 2016). In this case, the best-fit $N_{HI}$ value is a simple approximation of the overall HI  column densities. Interestingly, the best fit $N_{HI}$ of the five LyC leakers are about $10^{17-20}~cm^{-2}$, larger than the $N_{HI}$ that permit LyC escape. It suggests that their LyC emission probably escape through some holes in the interstellar medium with much lower $N_{HI}$.

\begin{figure*}[ht]
\centering
  \includegraphics[width=0.9\textwidth]{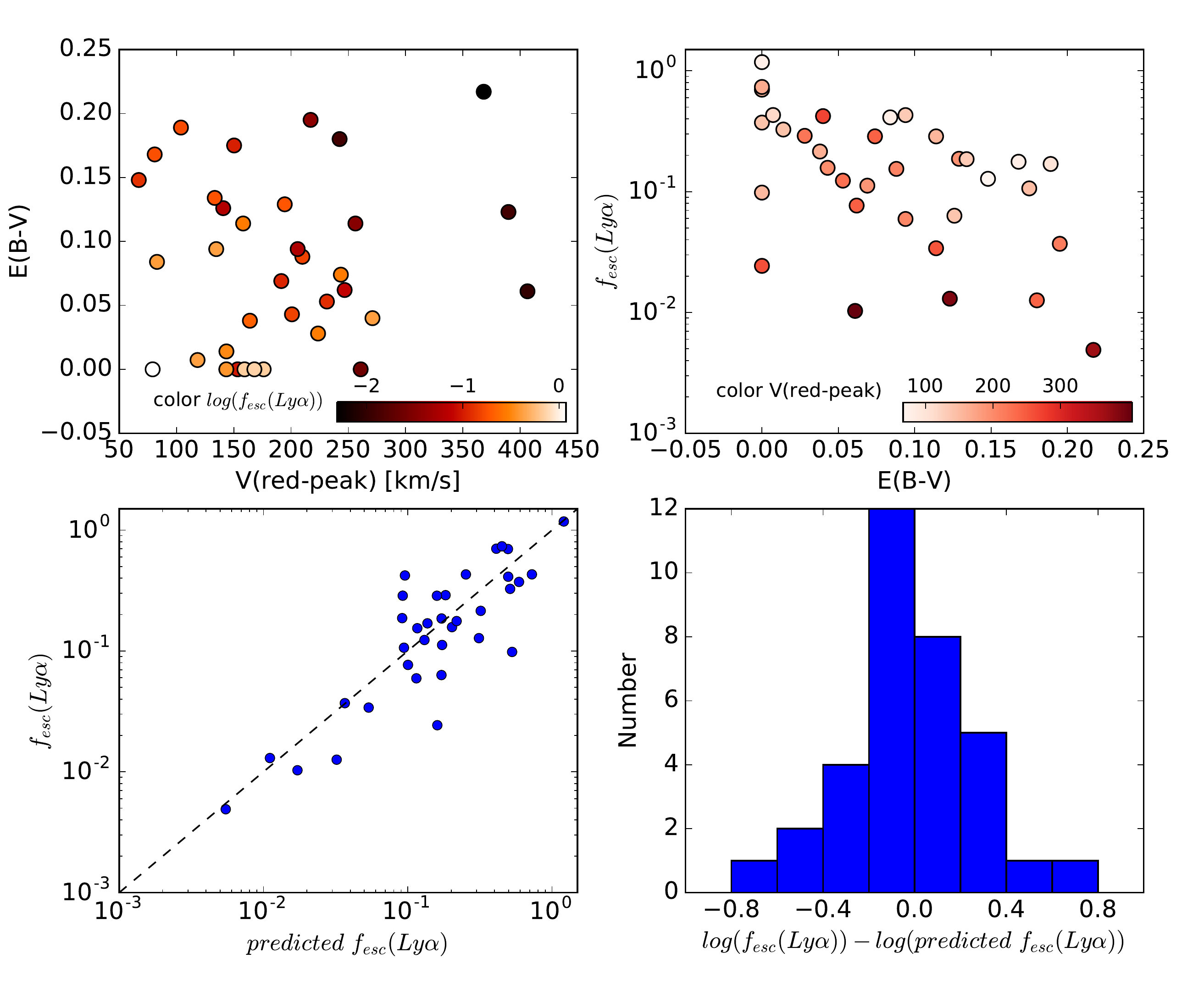}
 \caption{Top-left: the relation of E(B-V) vs. V(red-peak); The color-bar shows log(\fesc) value.  Top-right: the relation of \fesc\ and E(B-V); The color-bar shows V(red-peak) value. Bottom-left: the comparison of observed and predicted \fesc. Here the predicted log(\fesc)=$a\times(E(B-V)/0.1) + b\times(V(redpeak)/100) + c$. Bottom-right: the histogram of the differences, log(\fesc)-log(predicted \fesc).}
\end{figure*}

\section{Predicting \lya\ escape fraction}
As we said in the Introduction, one major reason for the studies of \lya\ escape is to use \lya\ lines to probe reionization. A fraction (\fesc) of intrinsic \lya\ photons first escape out of an LAE, then they go through the IGM where they can be further scattered by HI, and the remaining photons can finally be observed as a \lya\ line. So the IGM transmission can be measured from the observed \lya\ line flux if we know the intrinsic \lya\ line flux and \fesc, i.e. $IGM\ Transmission=(Observed~Ly\alpha)/(Intrinsic~Ly\alpha~\times~$\fesc). In the near future, JWST will be able to measure the observed \lya\ line and derive the intrinsic \lya\ line from the observed \ha\ line for galaxies in the epoch of reionization. If the remaining factor, \fesc, can be predicted from other observed galactic properties, then each \lya\ line can be used as an IGM probe on its line of sight. With this sample of Green Peas, we have found correlations between \fesc\ and \lya\ kinematic features, dust extinction, metallicity, stellar mass, and HI column density. So can we select a few observable factors and fit an empirical relation to predict \fesc? 

Physically, \lya\ escape depends on the properties of dust and HI gas, so we should select the factors that can indicate the properties of dust and HI gas. Dust extinction is relatively easy to measure and could be a useful factor.  The \lya\ kinematic features strongly depend on the column density and kinematics of HI gas and could be another useful factor. Among a few \lya\ kinematic features, the \lya\ red-peak velocity is easier and more robust to measure than the blue-peak velocity which might be removed by absorption and the line width which depends on the spectra resolution. The other three factors -- metallicity, stellar mass, and HI column density from fitting of \lya\ profile -- are difficult to measure and the uncertainties are large. Furthermore, both dust extinction and \lya\ V(red-peak) show relatively tight anti-correlations with \fesc. So we fit an linear empirical relation to predict \fesc\ from dust extinction and V(red-peak) of \lya\ profile. 

In figure 11, we first show the relations of \fesc, E(B-V), and V(red-peak). In the diagram of E(B-V) vs. V(red-peak), objects are color-coded by \fesc.  We can see that (i) E(B-V) and V(red-peak) don't show a correlation; (ii) the Green Peas with lower dust extinction and smaller V(red-peak) have larger \fesc.  In the diagram of \fesc\ vs. E(B-V), objects are color-coded by V(red-peak). Those Green Peas with large V(red-peak) generally have smaller \fesc\ than the others {\it with the same E(B-V)}.  Then we fit 37 Green Peas with both V(red-peak) and E(B-V) measurements. Two Green Peas, GP1454$+$4528 with gas inflow and GP0749$+$3337 with the largest V(red-peak), are outliers of the fitting, so we remove these two objects. The final best-fit relation of 35 Green Peas is 

\vspace{0.1cm}
\[ log(f^{Ly\alpha}_{esc})=a\times(E(B-V)/0.1) + b\times(V(redpeak)/100) + c\] 
\vspace{0.1cm}

, where $(a=-0.437,\  b=-0.483,\  c=0.464$). In the bottom two panels of figure 11, we compare the observed and the predicted \fesc\ and show the histogram of the differences, log(\fesc)-log(predicted \fesc). The standard deviation of this relation is 0.3 dex. 

Now we have a relation to predict \fesc\ from dust extinction and \lya\ V(red-peak). If JWST measures the observed \lya\ flux, observed \ha\ flux, dust extinction, and \lya\ V(red-peak) of a $z>7$ LAE, then we can infer the IGM transmission along this line of sight using the formula $IGM\ Transmission=(Observed~Ly\alpha)/(Intrinsic~Ly\alpha~\times~$\fesc), where the ``Intrinsic \lya" is calculated from dust extinction corrected \ha\ flux and \fesc\ is calculated from the empirical relation.

The IGM measured by this method is the ``true" IGM far from the LAE, which is in contrast to the circum-galactic medium (CGM). 
The ``true" IGM only affects the strength of \lya\ red peak by the damped absorption factor of $e^{-\tau}$, where $\tau$ is the optical depth of the IGM HI gas along the line of sight, and its effect on the velocity of the narrow \lya\ red peak is negligible. Some simulations suggested that the HI gas in the CGM can be very close to the \lya\ photons in frequency, so the CGM HI gas can resonantly scatter and/or absorb \lya\ photons at V(red-peak)$<$160 \kms\ (Laursen et al. 2011; Dijkstra 2014) and change the V(red-peak) of \lya\ profile. In fact those scatterings by CGM are part of the \lya\ escape process before \lya\ photons reach the ``true" IGM. So the influence of CGM gas is already considered in the empirical relation.

This empirical relation has important implications for reionization tests with \lya\ lines. Some observations suggested that the fraction of \lya\ emission line in Lyman-break galaxies drops rapidly at $z>6.5$ (e.g. Hayes et al. 2011; Tilvi et al. 2014; Pentericci et al. 2014). This could be due to small number statistics. But if this signal is real, it suggests either (i) the ``true" IGM  optical depth increases rapidly or (ii) the optical depth of ISM and CGM increases rapidly. Using our empirical relation, we can measure the optical depth of the ``true" IGM and distinguish these two possibilities.

Some recent observations suggest that five $z\sim7$ galaxies show very small velocity offsets about 20-150\kms\ between \lya\ and [CII] emission lines (Pentericci et al. 2016; Bradac et al. 2017). Those small V(red-peak) values may indicate that the \lya\ escape fractions are high and the optical depths of ISM and CGM are small. 

One caveat regards whether the empirical relation derived from low-$z$ analogs is applicable to high-$z$ LAEs.  The properties of ISM and CGM likely evolve between the low-$z$ LAEs (Green Peas) and LAEs in the epoch of reionization. 
However, since the physics of \lya\ resonant scattering is same in both low and high-$z$, increasing the HI gas column density in ISM probably doesn't change how $N_{HI}$ affects \lya\ profile. So the empirical relation is very likely applicable to $z>6$ LAEs.

\section{Conclusion}
We studied \lya\ escape in a statistical sample of Green Peas with HST/COS \lya\ spectra. About 2/3 Green Peas show strong \lya\ emission lines. Many Green Peas show double-peaked \lya\ line profiles, but the \lya\ profiles are diverse. These Green Peas have well measured galactic properties from SDSS optical spectra, so we investigated the dependence of \lya\ escape on dust extinction, metallicity, stellar mass, galaxy morphology, and [OIII]/[OII] ratio. We also fit their \lya\ profiles with the HI shell radiative transfer model. Finally, we derived an empirical relation to predict \lya\ escape fraction. Our major conclusions are as follows: 
 
\begin{enumerate}

\item With a statistical sample of 43 Green Peas that cover the whole ranges of dust extinction and metallicity properties of Green Peas, we found about 2/3 of Green Peas are strong \lya\ line emitters with distribution of EW(\lya) consistent with high-$z$ LAEs. This confirmed that Green Peas generally are the best analogs of high-$z$ LAEs in the nearby universe. 

\item The \fesc\ shows anti-correlations with a few \lya\ kinematic features -- the blue peak velocity, the red peak velocity, the peak separation, and the FWHM(red) of \lya\ profile. These \lya\ kinematic features are sensitive to the column density and the kinematics of HI gas. As more scatterings in HI gas can make the \lya\ velocity offsets larger and the \lya\ profile broader, these correlations strongly suggest low $N_{HI}$ and fewer scatterings help \lya\ photons escape. 

\item With a large sample, we found many {\em correlations} regarding the dependence of \lya\ escape on galactic properties -- \fesc\ generally increases at lower dust extinction, lower metallicity, lower stellar mass, and higher [OIII]/[OII] ratio. \fesc\ does not have an obvious relation with the UV morphology of Green Peas.

\item The single shell radiative transfer model can reproduce most \lya\ profiles of Green Peas. The best-fit $N_{HI}$ anti-correlates with \fesc, indicating that low $N_{HI}$ is key to \lya\ escape. 

\item We fit an empirical linear relation between \fesc, dust extinction, and \lya\ red peak velocity. This relation can be used to predict the \fesc\ of LAEs and isolate the effect of IGM scatterings from \lya\ escape. As JWST can measure the dust extinction and \lya\ red peak velocity of some $z > 7$ LAEs, this relation makes it possible to measure the HI column density of IGM along the line of sight of each LAE and to probe reionization with their \lya\ lines.

\end{enumerate}

\acknowledgments
We thank David Sobral, Edmund Christian Herenz, Kimihiko Nakajima, Alaina Henry, and the referee for very helpful comments. The imaging and spectroscopy data are based on observations with the NASA / ESA Hubble Space Telescope, obtained at the Space Telescope Science Institute, which is operated by the Association of Universities for Research in Astronomy (AURA), Inc., under NASA contract NAS 5-26555. Some of the data presented in this paper were obtained from the Mikulski Archive for Space Telescopes (MAST). STScI is operated by the Association of Universities for Research in Astronomy, Inc., under NASA contract NAS5-26555. Support for MAST for non-HST data is provided by the NASA Office of Space Science via grant NNX09AF08G and by other grants and contracts. H.Y. acknowledges support from China Scholarship Council. H.Y. and J.X.W. thanks supports from NSFC 11233002, 11421303,  and CAS Frontier Science Key Research Program (QYZDJ-SSW-SLH006). This work has also been supported in part by NSF grant AST-1518057; and by support for HST program \#14201.

\begin{deluxetable*}{llllccccccc}
\tablecaption{The Sample}
\tabletypesize{\scriptsize}
\centering

\tablehead{\colhead{ID} & \colhead{RA} & \colhead{DEC} & \colhead{\it{z}} &  \colhead{E(B-V)$_{MW}$} & \colhead{E(B-V)} & \colhead{12+log(O/H)}  &\colhead{log(M/M$_{\odot}$)} & \colhead{SFR}  &  \colhead{$R_{e}$}   &  \colhead{GO\#} \\
\colhead{(1)} & \colhead{(2)} & \colhead{(3)} & \colhead{(4)} & \colhead{(5)} & \colhead{(6)} & \colhead{(7)} & \colhead{(8)} &\colhead{(9)} &\colhead{(10)} & \colhead{(11)} } 

\startdata
1333$+$6246$^{a}$ & 13:33:03.94 & $+$62:46:03.7 & 0.31812 & 0.017 & 0.000 & 7.72 & 8.50 & 1.4 & 0.72 & 13744 \\
1559$+$0841 & 15:59:25.97 & $+$08:41:19.1 & 0.29704 & 0.033 & 0.000 & 8.04 & 8.97 & 3.5 & 0.47 & 14201 \\
1219$+$1526 & 12:19:03.98 & $+$15:26:08.5 & 0.19560 & 0.022 & 0.000 & 7.81 & 8.35 & 13.0 & 0.33 & 12928 \\
1514$+$3852 & 15:14:08.63 & $+$38:52:07.3 & 0.33262 & 0.019 & 0.000 & 8.12 & 9.32 & 6.4 & 0.67 & 14201 \\
1503$+$3644$^{a}$ & 15:03:42.82 & $+$36:44:50.8 & 0.35569 & 0.013 & 0.007 & 8.01 & 8.22 & 12.9 & 0.52 & 13744 \\
1442$-$0209$^{a}$ & 14:42:31.37 & $-$02:09:52.8 & 0.29367 & 0.046 & 0.094 & 7.95 & 8.96 & 21.2 & 0.50 & 13744 \\
1133$+$6514 & 11:33:03.80 & $+$65:13:41.3 & 0.24140 & 0.009 & 0.040 & 7.95 & 9.30 & 6.4 & 0.82 & 12928 \\
1249$+$1234 & 12:48:34.64 & $+$12:34:02.9 & 0.26339 & 0.026 & 0.084 & 8.10 & 9.05 & 18.3 & 0.71 & 12928 \\
1009$+$2916 & 10:09:18.99 & $+$29:16:21.5 & 0.22192 & 0.019 & 0.000 & 7.92 & 7.87 & 3.7 & 0.46 & 14201 \\
0815$+$2156 & 08:15:52.00 & $+$21:56:23.6 & 0.14095 & 0.035 & 0.014 & 7.96 & 8.71 & 4.4 & 0.35 & 13293 \\
1424$+$4217 & 14:24:05.73 & $+$42:16:46.3 & 0.18479 & 0.009 & 0.028 & 8.02 & 8.34 & 19.2 & 0.48 & 12928 \\
0926$+$4428 & 09:26:00.44 & $+$44:27:36.5 & 0.18069 & 0.016 & 0.074 & 8.02 & 8.78 & 14.8 & 0.43 & 11727 \\
1152$+$3400$^{a}$ & 11:52:04.88 & $+$34:00:49.8 & 0.34195 & 0.017 & 0.114 & 7.95 & 8.35 & 23.2 & 0.52 & 13744 \\
0021$+$0052 & 00:21:01.02 & $+$00:52:48.1 & 0.09836 & 0.021 & 0.038 & 8.14 & 9.30 & 13.7 & 0.44 & 13017 \\
1122$+$6154 & 11:22:19.73 & $+$61:54:45.4 & 0.20456 & 0.007 & 0.129 & 8.14 & 7.85 & 6.5 & 0.32 & 14201 \\
0925$+$1403$^{a}$ & 09:25:32.37 & $+$14:03:13.0 & 0.30121 & 0.027 & 0.134 & 8.01 & 8.46 & 23.8 & 0.42 & 13744 \\
0911$+$1831 & 09:11:13.34 & $+$18:31:08.2 & 0.26220 & 0.024 & 0.168 & 7.96 & 9.75 & 26.8 & 0.57 & 12928 \\
0917$+$3152 & 09:17:02.52 & $+$31:52:20.5 & 0.30036 & 0.017 & 0.189 & 8.10 & 9.37 & 21.8 & 0.47 & 14201 \\
1137$+$3524 & 11:37:22.14 & $+$35:24:26.7 & 0.19439 & 0.016 & 0.043 & 8.12 & 9.56 & 19.5 & 0.72 & 12928 \\
1025$+$3622 & 10:25:48.38 & $+$36:22:58.4 & 0.12649 & 0.010 & 0.088 & 8.11 & 9.20 & 10.0 & 0.76 & 13017 \\
1440$+$4619 & 14:40:09.94 & $+$46:19:36.9 & 0.30076 & 0.012 & 0.148 & 8.13 & 9.62 & 38.0 & 0.72 & 14201 \\
1429$+$0643 & 14:29:47.03 & $+$06:43:34.9 & 0.17351 & 0.022 & 0.053 & 8.01 & 9.40 & 30.6 & 0.40 & 13017 \\
1054$+$5238 & 10:53:30.83 & $+$52:37:52.9 & 0.25264 & 0.013 & 0.069 & 8.08 & 9.77 & 27.3 & 0.62 & 12928 \\
1428$+$1653 & 14:28:56.41 & $+$16:53:39.4 & 0.18164 & 0.017 & 0.175 & 8.12 & 9.60 & 22.2 & 0.77 & 13017 \\
0303$-$0759 & 03:03:21.41 & $-$07:59:23.2 & 0.16488 & 0.085 & 0.000 & 7.87 & 9.15 & 8.9 & 0.56 & 12928 \\
1244$+$0216 & 12:44:23.37 & $+$02:15:40.4 & 0.23943 & 0.021 & 0.062 & 8.09 & 9.65 & 31.0 & 1.02 & 12928 \\
2237$+$1336 & 22:37:35.05 & $+$13:36:47.0 & 0.29350 & 0.049 & 0.126 & 8.11 & 9.45 & 30.7 & 1.08 & 14201 \\
1454$+$4528 & 14:54:35.58 & $+$45:28:56.3 & 0.26851 & 0.036 & 0.169 & 8.22 & 9.52 & 21.4 & 0.45 & 14201 \\
1018$+$4106 & 10:18:03.24 & $+$41:06:21.0 & 0.23705 & 0.012 & 0.094 & 7.93 & 9.32 & 10.4 & 0.78 & 14201 \\
0751$+$1638 & 07:51:57.78 & $+$16:38:13.2 & 0.26471 & 0.031 & 0.149 & 7.85 & 8.35 & 7.8 & 0.80 & 14201 \\
0822$+$2241 & 08:22:47.66 & $+$22:41:44.0 & 0.21619 & 0.039 & 0.195 & 8.11 & 8.43 & 41.6 & 0.68 & 14201 \\
1339$+$1516 & 13:39:28.30 & $+$15:16:42.1 & 0.19202 & 0.026 & 0.114 & 8.05 & 9.43 & 18.7 & 0.38 & 14201 \\
1543$+$3446 & 15:43:01.22 & $+$34:46:01.4 & 0.18733 & 0.025 & 0.000 & 7.96 & 8.05 & 2.6 & 0.77 & 14201 \\
0938$+$5428 & 09:38:13.49 & $+$54:28:25.0 & 0.10208 & 0.015 & 0.123 & 8.17 & 9.40 & 13.6 & 0.47 & 11727 \\
0927$+$1740 & 09:27:28.67 & $+$17:40:18.6 & 0.28831 & 0.026 & 0.180 & 8.06 & 9.26 & 18.2 & 0.94 & 14201 \\
1457$+$2232 & 14:57:35.13 & $+$22:32:01.7 & 0.14861 & 0.041 & 0.061 & 8.02 & 9.13 & 11.6 & 0.42 & 13293 \\
0749$+$3337 & 07:49:36.77 & $+$33:37:16.3 & 0.27318 & 0.048 & 0.203 & 8.18 & 9.49 & 62.3 & 1.47 & 14201 \\
1032$+$2717 & 10:32:26.95 & $+$27:17:55.2 & 0.19246 & 0.018 & 0.097 & 8.22 & 9.65 & 13.3 & 0.63 & 14201 \\
0805$+$0925 & 08:05:18.04 & $+$09:25:33.5 & 0.33034 & 0.018 & 0.402 & 7.98 & 9.36 & 22.9 & 0.81 & 14201 \\
1205$+$2620 & 12:05:00.67 & $+$26:20:47.7 & 0.34261 & 0.016 & 0.178 & 7.89 & 9.84 & 22.0 & 0.83 & 14201 \\
0055$-$0021 & 00:55:27.46 & $-$00:21:48.7 & 0.16745 & 0.022 & 0.217 & 8.18 & 9.70 & 30.4 & 0.46 & 11727 \\
0339$-$0725 & 03:39:47.79 & $-$07:25:41.2 & 0.26071 & 0.053 & 0.095 & 8.31 & 9.70 & 29.6 & 0.88 & 14201 \\
0747$+$2336 & 07:47:58.00 & $+$23:36:32.7 & 0.15524 & 0.051 & 0.085 & 8.02 & 9.06 & 5.9 & 0.59 & 14201 
\enddata

\tablecomments{Column Descriptions: (1) Object ID; (4) Redshifts are from SDSS optical spectra; (5) The Milky Way extinction $E(B-V)_{MW}$, based on Schlafly \& Finkbeiner (2011); (6) dust extinction; (7) metallicity; (8) stellar mass; (9) star formation rate in unit of M$_{\odot}~yr^{-1}$ derived from \ha\ luminosity; (10) half light radius in unit of Kpc; (11) HST programs: GO14201 (PI S. Malhotra), GO13744 (PI T. Thuan; Izotov et al. 2016), GO13293 (PI A. Jaskot; Jaskot et al. 2014), GO12928 (PI A. Henry; Henry et al. 2015), GO11727 and GO13017 (PI T. Heckman; Heckman et al. 2011; Alexandroff et al. 2015).  These 43 galaxies are sorted by decreasing \fesc\ from top to bottom. The machine readable table is available online.}
\tablenotetext{a}{These are confirmed LyC leakers from Izotov et al. (2016).}

\end{deluxetable*}

\begin{deluxetable*}{lcccccccc}
\tablecaption{The line measurements from SDSS spectra}
\tabletypesize{\scriptsize}
\centering

\tablehead{\colhead{ID} & \colhead{[OII]3727} & \colhead{[OIII]4363} & \colhead{\hb} &  \colhead{[OIII]4959} & \colhead{[OIII]5007} & \colhead{\ha}  &\colhead{EW(\ha)} & \colhead{[OIII]/[OII]} \\
\colhead{(1)} & \colhead{(2)} & \colhead{(3)} & \colhead{(4)} & \colhead{(5)} & \colhead{(6)} & \colhead{(7)} & \colhead{(8)} &\colhead{(9)}  } 

\startdata
1333$+$6246 & 115$\pm$5 & 13.6$\pm$2.7 & 58$\pm$4 & 129$\pm$2 & 390$\pm$6 & 78$\pm$4 & 538 & 4.5 \\
1559$+$0841 & 169$\pm$4 & 9.4$\pm$1.3 & 124$\pm$12 & 229$\pm$2 & 693$\pm$6 & 232$\pm$11 & 288 & 5.5 \\
1219$+$1526 & 467$\pm$8 & 108.8$\pm$4.9 & 776$\pm$9 & 1635$\pm$8 & 4953$\pm$25 & 2207$\pm$18 & 744 & 14.2 \\
1514$+$3852 & 270$\pm$5 & 9.2$\pm$3.1 & 139$\pm$3 & 248$\pm$2 & 751$\pm$6 & 327$\pm$6 & 232 & 3.7 \\
1503$+$3644 & 220$\pm$3 & 19.3$\pm$1.6 & 184$\pm$2 & 397$\pm$2 & 1203$\pm$7 & 534$\pm$21 & 921 & 7.1 \\
1442$-$0209 & 248$\pm$5 & 32.2$\pm$2.4 & 299$\pm$4 & 647$\pm$5 & 1960$\pm$14 & 988$\pm$13 & 858 & 9.0 \\
1133$+$6514 & 268$\pm$5 & 19.2$\pm$2.3 & 196$\pm$3 & 376$\pm$2 & 1138$\pm$7 & 592$\pm$6 & 263 & 5.3 \\
1249$+$1234 & 575$\pm$8 & 28.8$\pm$2.4 & 364$\pm$5 & 740$\pm$5 & 2242$\pm$14 & 1169$\pm$12 & 717 & 4.6 \\
1009$+$2916 & 139$\pm$4 & 22.0$\pm$2.3 & 173$\pm$4 & 382$\pm$3 & 1156$\pm$9 & 473$\pm$16 & 422 & 11.1 \\
0815$+$2156 & 293$\pm$5 & 56.4$\pm$3.3 & 462$\pm$5 & 1065$\pm$7 & 3227$\pm$22 & 1383$\pm$13 & 717 & 14.0 \\
1424$+$4217 & 1129$\pm$16 & 114.9$\pm$3.6 & 1119$\pm$11 & 2463$\pm$11 & 7459$\pm$33 & 3333$\pm$25 & 629 & 8.4 \\
0926$+$4428 & 1090$\pm$14 & 56.3$\pm$3.5 & 733$\pm$8 & 1318$\pm$7 & 3994$\pm$22 & 2314$\pm$18 & 437 & 4.4 \\
1152$+$3400 & 237$\pm$4 & 22.5$\pm$1.3 & 228$\pm$3 & 461$\pm$2 & 1397$\pm$7 & 756$\pm$5 & 497 & 6.7 \\
0021$+$0052 & 5172$\pm$29 & 127.8$\pm$6.7 & 2909$\pm$14 & 4275$\pm$12 & 12949$\pm$36 & 8855$\pm$32 & 320 & 3.1 \\
1122$+$6154 & 257$\pm$5 & 11.6$\pm$1.3 & 200$\pm$4 & 369$\pm$3 & 1118$\pm$10 & 667$\pm$10 & 495 & 4.9 \\
0925$+$1403 & 297$\pm$7 & 25.8$\pm$4.0 & 282$\pm$4 & 596$\pm$4 & 1806$\pm$11 & 960$\pm$10 & 633 & 6.7 \\
0911$+$1831 & 576$\pm$10 & 15.5$\pm$3.2 & 379$\pm$5 & 442$\pm$3 & 1340$\pm$9 & 1343$\pm$14 & 348 & 2.5 \\
0917$+$3152 & 300$\pm$5 & 6.2$\pm$2.3 & 210$\pm$3 & 244$\pm$1 & 739$\pm$4 & 760$\pm$7 & 250 & 2.5 \\
1137$+$3524 & 1519$\pm$17 & 51.1$\pm$2.7 & 941$\pm$10 & 1563$\pm$7 & 4733$\pm$21 & 2865$\pm$21 & 434 & 3.9 \\
1025$+$3622 & 1816$\pm$17 & 60.7$\pm$4.5 & 1038$\pm$10 & 1746$\pm$10 & 5289$\pm$31 & 3318$\pm$25 & 312 & 3.4 \\
1440$+$4619 & 895$\pm$11 & 19.5$\pm$3.0 & 441$\pm$5 & 637$\pm$4 & 1929$\pm$12 & 1513$\pm$14 & 325 & 2.4 \\
1429$+$0643 & 2245$\pm$23 & 152.3$\pm$6.8 & 1785$\pm$15 & 3503$\pm$15 & 10610$\pm$46 & 5524$\pm$37 & 686 & 5.8 \\
1054$+$5238 & 1068$\pm$13 & 32.5$\pm$3.3 & 661$\pm$7 & 982$\pm$6 & 2974$\pm$17 & 2068$\pm$16 & 304 & 3.4 \\
1428$+$1653 & 1574$\pm$17 & 19.6$\pm$3.0 & 706$\pm$8 & 733$\pm$4 & 2220$\pm$13 & 2511$\pm$20 & 261 & 1.5 \\
0303$-$0759 & 488$\pm$8 & 74.0$\pm$2.5 & 656$\pm$7 & 1301$\pm$8 & 3941$\pm$23 & 1963$\pm$18 & 608 & 10.3 \\
1244$+$0216 & 1252$\pm$12 & 64.1$\pm$3.3 & 853$\pm$7 & 1681$\pm$8 & 5091$\pm$25 & 2665$\pm$18 & 667 & 4.9 \\
2237$+$1336 & 733$\pm$9 & 19.3$\pm$2.7 & 376$\pm$4 & 587$\pm$4 & 1780$\pm$11 & 1291$\pm$12 & 353 & 2.6 \\
1454$+$4528 & 498$\pm$8 & 9.2$\pm$3.2 & 293$\pm$4 & 401$\pm$3 & 1215$\pm$8 & 1033$\pm$11 & 277 & 2.6 \\
1018$+$4106 & 292$\pm$5 & 28.5$\pm$1.9 & 263$\pm$3 & 539$\pm$3 & 1633$\pm$10 & 846$\pm$8 & 570 & 6.5 \\
0751$+$1638 & 216$\pm$6 & 10.7$\pm$3.6 & 115$\pm$3 & 187$\pm$2 & 567$\pm$5 & 401$\pm$7 & 299 & 2.8 \\
0822$+$2241 & 1063$\pm$11 & 54.0$\pm$3.1 & 781$\pm$6 & 1551$\pm$6 & 4699$\pm$19 & 2886$\pm$17 & 605 & 4.4 \\
1339$+$1516 & 602$\pm$8 & 61.4$\pm$3.2 & 667$\pm$7 & 1485$\pm$6 & 4499$\pm$18 & 2222$\pm$57 & 523 & 8.4 \\
1543$+$3446 & 185$\pm$6 & 15.4$\pm$2.3 & 194$\pm$5 & 343$\pm$4 & 1037$\pm$11 & 480$\pm$7 & 342 & 7.5 \\
0339$-$0725 & 978$\pm$12 & 14.7$\pm$2.3 & 538$\pm$6 & 733$\pm$4 & 2220$\pm$13 & 1786$\pm$15 & 345 & 2.6 \\
0938$+$5428 & 3305$\pm$28 & 67.9$\pm$4.0 & 1887$\pm$15 & 2627$\pm$13 & 7957$\pm$39 & 6313$\pm$39 & 353 & 2.7 \\
0927$+$1740 & 328$\pm$6 & 17.7$\pm$2.1 & 196$\pm$4 & 419$\pm$3 & 1268$\pm$9 & 707$\pm$9 & 707 & 4.0 \\
1457$+$2232 & 764$\pm$9 & 103.1$\pm$2.7 & 868$\pm$12 & 2096$\pm$12 & 6349$\pm$36 & 2758$\pm$20 & 707 & 9.9 \\
0749$+$3337 & 1312$\pm$14 & 20.3$\pm$2.4 & 652$\pm$7 & 811$\pm$5 & 2457$\pm$14 & 2447$\pm$27 & 361 & 1.9 \\
1032$+$2717 & 845$\pm$8 & 25.1$\pm$0.8 & 520$\pm$5 & 910$\pm$4 & 2757$\pm$13 & 1687$\pm$13 & 651 & 3.8 \\
0805$+$0925 & 73$\pm$5 & 4.3$\pm$3.5 & 72$\pm$4 & 123$\pm$3 & 371$\pm$8 & 333$\pm$6 & 353 & 4.0 \\
1205$+$2620 & 324$\pm$5 & 9.0$\pm$3.7 & 165$\pm$3 & 198$\pm$1 & 599$\pm$4 & 587$\pm$8 & 350 & 1.9 \\
0055$-$0021 & 1738$\pm$11 & 29.0$\pm$3.0 & 956$\pm$5 & 1197$\pm$3 & 3626$\pm$10 & 3587$\pm$12 & 249 & 2.1 \\
0747$+$2336 & 363$\pm$5 & 35.8$\pm$1.8 & 354$\pm$5 & 796$\pm$5 & 2411$\pm$16 & 1166$\pm$11 & 366 & 7.6 
\enddata

\tablecomments{Observed line fluxes from SDSS spectra in units of $10^{-17}$ erg~s$^{-1}$~cm$^{-2}$. The EW(\ha) is rest-frame \ha\ equivalent width. The [OIII]/[OII] ratio are extinction corrected using the Calzetti et al. (2000) extinction law. The machine readable table is available online.}

\end{deluxetable*}

\begin{deluxetable*}{lccccccc}
\tablecaption{\lya\ properties}
\tabletypesize{\scriptsize}
\centering

\tablehead{\colhead{ID} & \colhead{\lya\ flux} & \colhead{log(L(\lya)~erg~s$^{-1}$)} & \colhead{EW(\lya)} &  \colhead{\fesc} & \colhead{V(blue-peak)} & \colhead{V(red-peak)}  &\colhead{FWHM(red)} \\
\colhead{} & \colhead{$10^{-16}$ erg~s$^{-1}$~cm$^{-2}$~} & \colhead{} & \colhead{\AA} & \colhead{} & \colhead{\kms} & \colhead{\kms} & \colhead{\kms}  \\
\colhead{(1)} & \colhead{(2)} & \colhead{(3)} & \colhead{(4)} & \colhead{(5)} & \colhead{(6)} & \colhead{(7)} & \colhead{(8)}  } 

\startdata
1333$+$6246$^{a}$ & 160.4$\pm$2.8 & 42.7 & 72.3 & 1.180 & $-^{d}$ & 79$\pm$17 & 245$\pm$24 \\
1559$+$0841 & 145.0$\pm$3.1 & 42.6 & 96.0 & 0.735 & $-$355$\pm$24 & 168$\pm$17 & 188$\pm$29 \\
1219$+$1526 & 1345.3$\pm$5.9 & 43.2 & 164.5 & 0.702 & $-$76$\pm$13 & 176$\pm$13 & 213$\pm$18 \\
1514$+$3852 & 180.8$\pm$4.1 & 42.8 & 60.0 & 0.698 & $-^{e}$ & 159$\pm$17 & 222$\pm$58 \\
1503$+$3644$^{a}$ & 195.2$\pm$4.3 & 42.9 & 106.6 & 0.431 & $-$349$\pm$43 & 118$\pm$23 & 229$\pm$27 \\
1442$-$0209$^{a}$ & 504.5$\pm$5.6 & 43.1 & 134.9 & 0.430 & $-^{d}$ & 135$\pm$26 & 267$\pm$25 \\
1133$+$6514 & 208.0$\pm$1.9 & 42.6 & 42.3 & 0.422 & $-$69$\pm$15 & 271$\pm$22 & 234$\pm$21 \\
1249$+$1234 & 528.0$\pm$2.6 & 43.1 & 101.8 & 0.412 & $-^{d}$ & 83$\pm$27 & 364$\pm$17 \\
1009$+$2916 & 142.8$\pm$2.5 & 42.3 & 69.5 & 0.373 & $-$116$\pm$45 & 144$\pm$22 & 206$\pm$26 \\
0815$+$2156 & 401.2$\pm$1.4 & 42.3 & 82.2 & 0.327 & $-$121$\pm$13 & 144$\pm$13 & 216$\pm$19 \\
1424$+$4217 & 858.6$\pm$4.1 & 42.9 & 89.5 & 0.290 & $-$150$\pm$32 & 224$\pm$10 & 208$\pm$16 \\
0926$+$4428 & 636.8$\pm$2.3 & 42.8 & 47.8 & 0.287 & $-$165$\pm$51 & 244$\pm$17 & 327$\pm$18 \\
1152$+$3400$^{a}$ & 248.6$\pm$4.6 & 43.0 & 74.5 & 0.287 & $-$146$\pm$26 & 158$\pm$44 & 235$\pm$30 \\
0021$+$0052 & 1523.5$\pm$9.7 & 42.6 & 32.8 & 0.215 & $-$418$\pm$38 & 164$\pm$12 & 253$\pm$16 \\
1122$+$6154 & 144.1$\pm$2.1 & 42.2 & 60.0 & 0.187 & $-$56$\pm$21 & 194$\pm$26 & 202$\pm$25 \\
0925$+$1403$^{a}$ & 225.1$\pm$4.1 & 42.8 & 90.0 & 0.186 & $-$145$\pm$24 & 133$\pm$23 & 160$\pm$25 \\
0911$+$1831 & 315.7$\pm$2.1 & 42.8 & 56.5 & 0.177 & $-$278$\pm$17 & 81$\pm$12 & 207$\pm$17 \\
0917$+$3152 & 167.7$\pm$3.3 & 42.7 & 38.0 & 0.169 & $-$209$\pm$42 & 104$\pm$22 & 189$\pm$25 \\
1137$+$3524 & 381.1$\pm$3.4 & 42.6 & 40.4 & 0.158 & $-$355$\pm$46 & 201$\pm$22 & 285$\pm$20 \\
1025$+$3622 & 436.6$\pm$3.6 & 42.3 & 26.3 & 0.154 & $-$248$\pm$39 & 210$\pm$12 & 253$\pm$15 \\
1440$+$4619 & 214.2$\pm$3.6 & 42.8 & 33.8 & 0.128 & $-^{d}$ & 67$\pm$29 & 248$\pm$30 \\
1429$+$0643 & 607.1$\pm$2.9 & 42.7 & 42.7 & 0.123 & $-$257$\pm$34 & 231$\pm$19 & 324$\pm$19 \\
1054$+$5238 & 153.5$\pm$2.6 & 42.5 & 17.7 & 0.112 & $-$314$\pm$88 & 192$\pm$12 & 225$\pm$25 \\
1428$+$1653 & 311.9$\pm$2.2 & 42.5 & 29.1 & 0.106 & $-$360$\pm$25 & 150$\pm$20 & 242$\pm$18 \\
0303$-$0759 & 99.6$\pm$2.1 & 41.9 & 14.2 & 0.098 & $-$313$\pm$48 & 153$\pm$26 & 270$\pm$23 \\
1244$+$0216 & 189.9$\pm$1.6 & 42.5 & 47.0 & 0.077 & $-$240$\pm$14 & 247$\pm$14 & 302$\pm$25 \\
2237$+$1336 & 51.4$\pm$2.6 & 42.1 & 15.3 & 0.063 & $-^{d}$ & 141$\pm$36 & 272$\pm$35 \\
1454$+$4528 & 72.3$\pm$2.1 & 42.2 & 30.0 & 0.061 & $-$56$\pm$58 & 444$\pm$18 & 202$\pm$47 \\
1018$+$4106 & 47.0$\pm$1.5 & 41.9 & 33.1 & 0.059 & $-$306$\pm$44 & 206$\pm$25 & 238$\pm$32 \\
0751$+$1638 & 13.9$\pm$1.3 & 41.5 & 15.8 & 0.043 & $-^{e}$ & $-^{e}$ & $-^{e}$ \\
0822$+$2241 & 156.5$\pm$2.9 & 42.3 & 51.6 & 0.037 & $-$304$\pm$65 & 217$\pm$31 & 262$\pm$39 \\
1339$+$1516 & 82.5$\pm$1.9 & 41.9 & 44.7 & 0.034 & $-$351$\pm$30 & 256$\pm$42 & 435$\pm$62 \\
1543$+$3446$^{b}$ & 10.6$\pm$0.8 & 41.0 & 5.4 & 0.024 & $-^{e}$ & 261$\pm$94 & 407$\pm$90 \\
0938$+$5428$^{b}$ & 107.1$\pm$2.0 & 41.5 & 3.5 & 0.013 & $-$279$\pm$20 & 390$\pm$17 & 333$\pm$43 \\
0927$+$1740$^{b}$ & 14.0$\pm$1.1 & 41.6 & 7.2 & 0.013 & $-^{e}$ & 242$\pm$91 & 408$\pm$150 \\
1457$+$2232$^{b}$ & 32.3$\pm$0.6 & 41.3 & 5.3 & 0.010 & $-$329$\pm$37 & 406$\pm$14 & 321$\pm$54 \\
0749$+$3337 & 9.2$\pm$1.7 & 41.3 & 8.9 & 0.010 & $-$427$\pm$72 & 568$\pm$72 & 405$\pm$114 \\
1032$+$2717$^{b}$ & 19.2$\pm$0.9 & 41.3 & 5.5 & 0.009 & $-^{e}$ & $-^{e}$ & $-^{e}$ \\
0805$+$0925$^{b}$ & 9.5$\pm$1.3 & 41.5 & 9.2 & 0.009 & $-^{e}$ & $-^{e}$ & $-^{e}$ \\
1205$+$2620$^{b}$ & 5.8$\pm$1.3 & 41.4 & 3.0 & 0.006 & $-^{e}$ & $-^{e}$ & $-^{e}$ \\
0055$-$0021$^{b}$ & 31.3$\pm$1.0 & 41.4 & 3.2 & 0.005 & $-^{e}$ & 368$\pm$45 & 420$\pm$39 \\
0339$-$0725$^{c}$ & $-$1.4$\pm$1.8 & $-$ & $-$ & $-$ & $-$ & $-$ & $-$ \\
0747$+$2336$^{c}$ & $-$ & $-$ & $-$ & $-$ & $-$ & $-$ & $-$ 
\enddata

\tablecomments{Column Descriptions: (1) Object ID; (2) \lya\ emission line flux; (3) \lya\ emission line luminosity; (4) equivalent width of \lya\ line; (5) \lya\ escape fraction; (6) Velocity of \lya\ blue peak; (7) Velocity of \lya\ red peak; (8) FWHM of the red portion of \lya\ profile.  These 43 galaxies are sorted by decreasing \fesc\ from top to bottom. The machine readable table is available online.}
\tablenotetext{a}{These are confirmed LyC leakers from Izotov et al. (2016).}
\tablenotetext{b}{These Green Peas show damped \lya\ absorption wings in their \lya\ spectra.}
\tablenotetext{c}{No \lya\ emission line was detected.}
\tablenotetext{d}{Their \lya\ profiles don't have blue peaks.}
\tablenotetext{e}{Their \lya\ profiles are too noisy for measuring \lya\ kinematic features.}

\end{deluxetable*}

\begin{deluxetable*}{cccccc}
\tablecaption{\LyA\ profile Model Parameters}
\tablehead{\colhead{ID} & \colhead{log($N_{HI}~cm^{-2}$)} & \colhead{$V_{exp}$} & \colhead{log(T)} & \colhead{$\tau_{d}$} & \colhead{$\sigma$} \\ 
\colhead{} & \colhead{} & \colhead{(\kms)} & \colhead{(K)} & \colhead{} & \colhead{\kms} \\
\colhead{(1)} & \colhead{(2)} & \colhead{(3)} & \colhead{(4)} & \colhead{(5)} & \colhead{(6)} } 
\startdata
1333$+$6246 & 19.39$^{+0.08}_{-0.07}$ & 270$^{+4}_{-4}$ & 5.0$^{+0.2}_{-0.1}$ & 0.71$^{+0.09}_{-0.08}$ & 125$^{+2}_{-2}$ \\
1559$+$0841 & 19.40$^{+0.07}_{-0.07}$ & 90$^{+3}_{-4}$ & 3.0$^{+0.1}_{-0.1}$ & 0.64$^{+0.08}_{-0.06}$ & 203$^{+2}_{-2}$ \\
1514$+$3852 & 19.20$^{+0.07}_{-0.07}$ & 80$^{+4}_{-3}$ & 3.8$^{+0.1}_{-0.1}$ & 0.01$^{+0.01}_{-0.00}$ & 305$^{+7}_{-7}$ \\
1503$+$3644 & 16.81$^{+0.08}_{-0.08}$ & 140$^{+4}_{-4}$ & 5.4$^{+0.2}_{-0.1}$ & 0.14$^{+0.07}_{-0.05}$ & 266$^{+5}_{-5}$ \\
1442$-$0209 & 18.80$^{+0.07}_{-0.07}$ & 150$^{+4}_{-4}$ & 4.2$^{+0.2}_{-0.2}$ & 0.01$^{+0.01}_{-0.00}$ & 230$^{+2}_{-2}$ \\
1009$+$2916 & 19.60$^{+0.07}_{-0.07}$ & 30$^{+4}_{-4}$ & 3.4$^{+0.1}_{-0.1}$ & 0.00$^{+0.00}_{-0.00}$ & 201$^{+3}_{-3}$ \\
1152$+$3400 & 20.00$^{+0.07}_{-0.07}$ & 5$^{+1}_{-1}$ & 3.4$^{+0.2}_{-0.1}$ & 0.01$^{+0.00}_{-0.00}$ & 333$^{+6}_{-6}$ \\
0021$+$0052 & 19.59$^{+0.07}_{-0.07}$ & 130$^{+4}_{-3}$ & 5.0$^{+0.1}_{-0.1}$ & 0.22$^{+0.01}_{-0.01}$ & 100$^{+2}_{-2}$ \\
1122$+$6154 & 19.98$^{+0.08}_{-1.85}$ & 7$^{+1}_{-26}$ & 3.4$^{+0.2}_{-0.2}$ & 0.00$^{+1.13}_{-0.00}$ & 259$^{+4}_{-5}$ \\
0925$+$1403 & 19.79$^{+0.07}_{-0.08}$ & 8$^{+1}_{-1}$ & 3.0$^{+0.2}_{-0.2}$ & 0.03$^{+0.00}_{-0.00}$ & 229$^{+5}_{-5}$ \\
0917$+$3152 & 19.00$^{+0.07}_{-0.07}$ & 60$^{+3}_{-3}$ & 3.5$^{+0.3}_{-0.2}$ & 0.01$^{+0.01}_{-0.01}$ & 275$^{+8}_{-8}$ \\
1025$+$3622 & 19.40$^{+0.07}_{-0.07}$ & 110$^{+3}_{-4}$ & 4.1$^{+0.2}_{-0.3}$ & 0.00$^{+0.01}_{-0.00}$ & 228$^{+7}_{-5}$ \\
1440$+$4619 & 19.18$^{+0.08}_{-0.08}$ & 259$^{+4}_{-4}$ & 5.0$^{+0.1}_{-0.1}$ & 0.23$^{+0.03}_{-0.03}$ & 117$^{+2}_{-2}$ \\
1429$+$0643 & 20.39$^{+0.08}_{-0.08}$ & 15$^{+2}_{-2}$ & 3.4$^{+0.1}_{-0.1}$ & 0.00$^{+0.00}_{-0.00}$ & 392$^{+3}_{-3}$ \\
1428$+$1653 & 16.04$^{+1.55}_{-0.09}$ & 168$^{+5}_{-11}$ & 5.7$^{+0.2}_{-0.4}$ & 0.10$^{+0.02}_{-0.09}$ & 205$^{+20}_{-4}$ \\
2237$+$1336 & 19.88$^{+0.15}_{-0.13}$ & 258$^{+44}_{-10}$ & 4.9$^{+0.2}_{-1.1}$ & 1.14$^{+0.86}_{-0.33}$ & 140$^{+12}_{-8}$ \\
1454$+$4528 & 16.44$^{+0.36}_{-0.09}$ & -171$^{+4}_{-4}$ & 5.4$^{+0.1}_{-0.1}$ & 4.78$^{+0.17}_{-0.34}$ & 427$^{+10}_{-8}$ \\
1018$+$4106 & 19.60$^{+0.07}_{-0.07}$ & 49$^{+4}_{-4}$ & 3.8$^{+0.2}_{-0.2}$ & 0.07$^{+0.04}_{-0.04}$ & 268$^{+12}_{-11}$ \\
0751$+$1638 & 19.39$^{+0.07}_{-0.08}$ & 121$^{+17}_{-20}$ & 4.4$^{+0.3}_{-1.3}$ & 0.17$^{+0.34}_{-0.13}$ & 326$^{+31}_{-29}$ \\
0822$+$2241 & 20.54$^{+0.11}_{-0.18}$ & 6$^{+3}_{-2}$ & 3.0$^{+0.1}_{-0.2}$ & 0.01$^{+0.02}_{-0.00}$ & 363$^{+10}_{-19}$ \\
1339$+$1516 & 19.00$^{+0.07}_{-0.07}$ & 100$^{+4}_{-4}$ & 3.1$^{+0.6}_{-0.2}$ & 4.86$^{+0.11}_{-0.23}$ & 345$^{+5}_{-5}$ \\
0938$+$5428 & 20.60$^{+0.08}_{-0.08}$ & 15$^{+2}_{-2}$ & 4.6$^{+0.2}_{-0.2}$ & 0.08$^{+0.01}_{-0.01}$ & 64$^{+15}_{-10}$ \\
0055$-$0021 & 20.40$^{+0.07}_{-0.07}$ & 60$^{+3}_{-4}$ & 4.2$^{+0.1}_{-0.1}$ & 0.07$^{+0.02}_{-0.02}$ & 322$^{+17}_{-17}$  
\enddata
\tablecomments{Column Descriptions: (2) HI column density of the outflowing HI shell; (3) outflowing velocity of the HI shell; (4) HI gas temperature including turbulent motion as well as the true temperature; (5) dust optical depth; (6) 1$\sigma$ width of the Gaussian profile of the intrinsic \LyA\ line. These 23 galaxies are sorted by decreasing \fesc\ from top to bottom. }
\end{deluxetable*}

\end{document}